%% file: main.tex
\documentclass[twocolumn]{aastex631}
\usepackage{amsmath}
\usepackage{enumitem}

\received{2022-05-03}
\revised{2022-06-27}
\accepted{2022-06-28}
\published{2022-08-24}

\submitjournal{ApJ}

\newenvironment{changemargin}[2]{%
\begin{list}{}{%
\setlength{\topsep}{0pt}%
\setlength{\textheight}{#1}%
\setlength{\topmargin}{#2}%
\setlength{\listparindent}{\parindent}%
\setlength{\itemindent}{\parindent}%
\setlength{\parsep}{\parskip}%
}%
\item[]}{\end{list}}

\newcommand{\ZenodoLink}{\href{https://doi.org/10.5281/zenodo.6504276}{DOI 10.5281/zenodo.6504276}} 
\newcommand{\augerwebsite}{\href{https://www.auger.org/science/public-data/data}{website}}
\graphicspath{{./}}

\begin{document}

\title{Arrival Directions of Cosmic Rays above 32 EeV from Phase One of the Pierre Auger Observatory}
\collaboration{0}{The Pierre Auger Collaboration}
\email{spokespersons@auger.org}

\begin{abstract}
A promising energy range to look for angular correlation between cosmic rays of extragalactic origin and their sources is at the highest energies, above few tens of EeV ($1\:{\rm EeV}\equiv 10^{18}\:$eV). Despite the flux of these particles being extremely low, the area of  ${\sim}\:3{,}000 \: \text{km}^2$ covered at the Pierre Auger Observatory, and the 17-year data-taking period of the Phase~1 of its operations, have enabled us to measure the arrival directions of more than 2,600 ultra-high energy cosmic rays above $32\:\text{EeV}$. We publish this data set, the largest available at such energies from an integrated exposure of $122{,}000 \: \text{km}^2\:\text{sr}\:\text{yr}$, and search it for anisotropies over the $3.4\pi$ steradians covered with the Observatory. Evidence for a deviation in excess of isotropy at intermediate angular scale, with ${\sim}\:15^\circ$ Gaussian spread or ${\sim}\:25^\circ$ top-hat radius, is obtained at the $4\:\sigma$ significance level for cosmic-ray energies above ${\sim}\:40\:\text{EeV}$.
\end{abstract}

\keywords{Ultra-high-energy cosmic radiation (1733), Cosmic ray astronomy (324), Clustering (1908), Active galaxies (17), Starburst galaxies (1570)}

\section{Introduction} \label{sec:intro}
Cosmic rays are observed up to the astounding energies of more than $10^{20}\:$eV, making them the most energetic particles known in the Universe. However, the origin of these particles remains elusive. The search for the sources of ultra-high energy cosmic rays (UHECRs), at energies above a few EeV ($1\:\text{EeV} \equiv 10^{18}\:\text {eV}$), is challenging since they are almost all charged particles and thus deflected by the magnetic fields permeating the interstellar, intra-halo and intergalactic media \citep[see e.g.][for an overview]{2019FrASS...6...23B}. These magnetic fields are difficult to study and their modeling is far from being complete. However, above a few tens of EeV, the deflections could be small enough for cosmic rays to retain some directional information on the position of their sources, at least for nuclei with a sufficiently small charge \citep[e.g.][]{2016APh....85...54E, 2019JCAP...05..004F}.

The cosmological volume within which UHECR sources should be sought is fortunately limited. Cosmic rays at EeV energies can interact with the photon backgrounds populating intergalactic space, through the so-called GZK effect  \citep{1966PhRvL..16..748G,1966JETPL...4...78Z}. In particular, protons are expected to undergo photo-pion production and nuclei photodissociation interactions. The mean free path for energy losses depends on the cosmic-ray mass and energy. At 100\:EeV, the loss length is of the order of $200-300\:$Mpc for proton and iron and $3-6\:$Mpc for intermediate nuclei such as helium and nitrogen (\citealt{2012APh....39...33A}; see also Figure~6 from \citealt{2021arXiv211105659A} for a recent overview). Such short distances mean that the sources of the highest-energy cosmic rays must be in the local universe.

The recent detection by the Pierre Auger Collaboration of a dipolar anisotropy in the arrival directions of UHECRs with energies above $8\:$EeV is evidence that the majority of UHECR sources are not in the Milky Way \citep{Science2017}. The direction of the dipole points ${\sim}\:120^\circ$ away from the Galactic center and is instead consistent at the $2\:\sigma$ level with the local distribution of stellar mass \citep[2MASS redshift survey,][]{2012ApJS..199...26H}, after accounting for the deflections expected in the Galactic magnetic field \citep{2012ApJ...761L..11J}. Even without relying on magnetic deflections, the case for a density of UHECR sources following local extragalactic structures is further strengthened by the consistency at the $1\:\sigma$ confidence level (C.L.) between the directions of the UHECR anti-dipole and of the Local Void at equatorial coordinates $(\alpha, \delta) = (294^\circ, 15^\circ)$ or Galactic coordinates $(l, b) = (51^\circ, -3^\circ)$ \citep{2021ApJS..256...15B}. Combined with the growth of the dipole amplitude with energy expected from the shrinking horizon out to which extragalactic sources remain visible \citep{2018ApJ...868....4A}, the properties of the large-scale anisotropy discovered by the Pierre Auger Collaboration provide a growing body of evidence against a Galactic origin of these cosmic rays. Which (classes of) extragalactic sources host UHECR accelerators nonetheless remains an open question.

In this article, we update previous searches for anisotropies at the highest energies \citep{2015ApJ...804...15A, 2018ApJ...853L..29A} with an unprecedentedly large data set. In particular, we exploit the entire \textit{Phase~1} of the Pierre Auger Observatory, i.e.\ the phase preceding the AugerPrime upgrade \citep{2016arXiv160403637T}. Important progress has been made on estimating the mass distribution of UHECRs using only the surface detector of the Observatory with its full duty cycle \citep[see e.g.][]{PhysRevD.93.072006,PierreAuger:2017tlx,AVE201723,DNN_Xmax_Aab_2021,DNN_Mu_Aab_2021}. However, the proposed methods are still not ready to be employed in arrival-direction studies, e.g.\ by selecting only the candidate light nuclei which would be less deflected by magnetic fields, should such a subsample exist in the data set. In the following, we then consider, as in previous works, only the energy and arrival direction of each event recorded with the Pierre Auger Observatory over 17 years of operation. 

The data set includes more than 2,600 events with energies $E\geq32\: \text{EeV}$ and zenith angles up to $80^\circ$, as described in Section~\ref{sec:data}. The release of this data set complements the publication of the arrival directions of events at energies between 4 and 8\:EeV and above 8\:EeV made available in \cite{Science2017}.\footnote{\url{https://www.auger.org/document-centre-public/download/78-data/4642-arrival-directions-8eev-science-2017}} The choice of an energy threshold at 32\:EeV for the present release anticipates upcoming publications focused on lower energy bins, namely $8-16\:$EeV and $16-32\:$EeV, as investigated e.g.\ in \cite{2018ApJ...868....4A} and \cite{2020ApJ...891..142A} where ${\sim}\:1{,}500$ and ${\sim}\:2{,}000$ events were studied above 32\:EeV, respectively. In Section~\ref{sec:blind}, we describe a first set of analyses that are not based on specific source models, i.e.\ a blind search for excesses in the sky, an autocorrelation study and the search for correlations with the Galactic and supergalactic planes as well as the Galactic center. Section~\ref{sec:likelihood} is devoted to the comparison of UHECR arrival directions with the expected flux pattern from specific classes of galaxies traced by their electromagnetic emission, from radio wavelengths to gamma rays. Finally, Section~\ref{sec:Cen} is devoted to a more in-depth study of the Centaurus region, which has intrigued the UHECR community since the early days of the Pierre Auger Observatory \citep{2007Science.1151124}.

To encourage further studies of the Phase~1 high-energy data set, this article is accompanied with supplementary materials. These include the data set itself in Appendix~\ref{app:data} and the dedicated analysis software in Appendix~\ref{app:code}. Appendix \ref{app:catalogs} describes the catalogs of galaxies used here.

\section{The data set} \label{sec:data}
The Pierre Auger Observatory \citep{AugerNIM2015} is located in Argentina near the town of Malarg\"ue. Stable data acquisition began on 1~January 2004. The Observatory is composed of a surface detector (SD) made of 1,660 water-Cherenkov stations distributed on a triangular grid overlooked with a fluorescence detector (FD). The FD consists of 27 telescopes at four locations on the perimeter of the SD array.

Here, we analyse the events with reconstructed energies larger than $32\:$EeV recorded with the SD array from 1~January 2004 to 31~December 2020. The SD is used to sample secondary particles in air showers and has full efficiency above 4\:EeV with ${\sim}\:100 \%$ duty cycle. 

Events recorded with SD are reconstructed differently based on their arrival direction in local coordinates: events with zenith angles, $\theta$, less than $60^\circ$ are called \textit{vertical} events, while events arriving with zenith angles from  $60^\circ$ to  $80^\circ$ are called \textit{inclined} events. Vertical events are included when the SD station with the largest signal is surrounded by at least four active stations. This \textit{a priori} condition is complemented by the \textit{a posteriori} requirement that the reconstructed core of the shower falls within an elementary isosceles triangle of active stations. These requirements ensure that the footprint of the shower is well-contained within the array, with ample data for an accurate reconstruction \citep{ABRAHAM201029}. Inclined events, on the other hand, are selected if the station closest to the reconstructed core position is surrounded by at least five active stations. Note that other analyses performed by the Pierre Auger Collaboration at lower energies may use a tighter selection. For example, the UHECR spectrum in \cite{2020PhRvD.102f2005A} is measured by requiring that all six active stations around the one with the highest signal are active. We are able to use a relaxed selection as the high-energy events included here all have large footprints, with an average of 17.7 triggered stations. We inspected each event and verified that the reconstruction was robust even with inactive stations in the core region. With respect to previous analyses, the identification of active stations that were not triggered has been improved to ensure a better selection. This was done through an \textit{a posteriori} check of the consistency of the signal distribution at ground: if a station is not triggered in a region of the array where the signal is more than twice that of the full trigger efficiency, which occurs for 11 events in the data set, the station is classified as non-active at the moment of the event \citep{ABRAHAM201029}.

The selection results in 2,040 events with $\theta < 60^\circ$ and 595 with $\theta \geq 60^\circ$ above $32\:\text{EeV}$.\footnote{To avoid border effects at the zenith angle separating the inclined and vertical selections, we identified events in the $60^\circ<\theta<62^\circ$ region that are well-reconstructed with the vertical procedure but not included in the inclined data set and, vice-versa, events in the $58^\circ<\theta<60^\circ$ region that are well-reconstructed with the inclined procedure but not included in the vertical data set. We found one event in the former case and none in the latter. } The exposure can be computed in a geometrical way since we are operating above the energy threshold for full efficiency for both data samples ($3\:$EeV for vertical and $4\:$EeV for inclined). The geometrical exposure for the selection and time span considered is $95{,}700 \: \text{km}^2 \: \text{sr} \:\text{yr}$ for the vertical sample and $26{,}300 \: \text{km}^2 \: \text{sr} \:\text{yr}$ for the inclined data set.

The reconstruction procedure for vertical events is described in detail in \cite{2020JInst..15P0021A}. The arrival direction is determined by fitting a spherical model to the arrival times of particles comprising the shower front. For inclined events, the reconstruction procedure is described in \cite{2014_JCAP_Inclined}. The arrival direction is, in this case, obtained by fitting the arrival times with a front which takes into account the muon propagation from its production point. For both data sets, the angular resolution, defined as the 68\% containment radius, is better than $1^\circ$ at all energies considered here.

The energy estimate is based on different observables for the two samples. The signal at a reference distance of $1000 \:$m from the shower core, $S(1000)$, is used for the vertical sample. The inclined reconstruction uses as estimator $N_{19}$, which represents the muon content of the shower with respect to a reference simulated proton shower with energy $E=10^{19}\:$eV. For both samples, a correction is applied to take into account the absorption that showers undergo at different zenith angles. This correction is performed through a data-driven procedure called \textit{constant intensity cut}, which is described in \cite{2020PhRvD.102f2005A}. The constant-intensity-cut method is used to convert $S(1000)$ and $N_{19}$ for each shower to the value they would have if the same shower had arrived from a reference zenith angle of $38^\circ$ and $68^\circ$ for vertical and inclined events, respectively. The corrected energy estimators, $S_{38}$ and $N_{68}$, are then calibrated using hybrid events, i.e.\ events observed with both the FD and the SD. Since the FD analysis enables a quasi-calorimetric measurement of the shower energy, the calibration procedure results in a reliable energy estimation for the whole SD data set without using air-shower simulations. The systematic uncertainty in the energy calibration is ${\sim}\:14\%$ while the energy resolution for the SD at the energies considered here is ${\sim}\:7\%$ \citep{2014_JCAP_Inclined,PhysRevLett.125.121106}.

We checked the consistency between the vertical and inclined data sets by comparing the ratio of number of events in the two samples, $N_{\rm incl}/N_{\rm vert}=0.292 \pm 0.014$, and the value expected from the ratio of geometrical exposures, accounting for the finite energy resolution of each data stream, $\frac{\omega_\text{incl}/c_\text{incl}({\ge}32\,\mathrm{EeV})}{\omega_\text{vert}/c_\text{vert}({\ge}32\,\mathrm{EeV})}=0.278$. In the latter ratio, $\omega$ is the geometrical exposure for each data set, which does not depend on energy, and $c({\geq}\:32\:\text{EeV})$ accounts for the net spillover of events from low to higher energies \citep[see the \textit{unfolding} procedure described in][]{2020PhRvD.102f2005A}. The ratios are in agreement at the $1\:\sigma$ C.L., showing that the vertical and inclined samples can be used together. To keep the analysis as data-driven as possible, we use the ratio of events observed above $32\:\text{EeV}$ as the expected exposure ratio when constructing simulated data sets above any energy threshold. It should be noted that at the highest energies probed here, $E\geq80\:\text{EeV}$, a deficit of inclined events is observed at a significance level of $2.5\:\sigma$. A further discussion of this deficit, which does not affect the results presented below, is provided in Appendix~\ref{app:data} together with the information on how to access the data.

\section{Search for overdensities and correlation with structures}
\label{sec:blind}
An earlier wide-ranging search with the Observatory for small- and intermediate-scale anisotropy was reported in \cite{2015ApJ...804...15A}. Searches for localized excesses in top-hat windows of angular radius $\Psi$ across the entire field of view of the Observatory, or around the Galactic center, Centaurus A and candidate host galaxies identified in multi-wavelength surveys, were performed by comparing the expected and observed numbers of events within the window. Similar analyses were performed along the Galactic and supergalactic planes, by counting the number of events within an angle $\Psi$ from these structures, and an autocorrelation study exploited the number of pairs of events separated by less than $\Psi$. The analyses were repeated above energy thresholds ranging from 40 to $80\:\text{EeV}$. An additional scan on the maximum distance of the sources was performed for analyses against catalogs of candidate host galaxies. Both scans in energy threshold and maximum distance were motivated by the limited horizon from which UHECR can reach Earth, although the determination of its observational value remains hindered by uncertainties on UHECR composition.

In this Section, we update the results presented in \cite{2015ApJ...804...15A}, with the exception of the search for correlation with catalogs, which is performed in Section~\ref{sec:likelihood}. 

\subsection{Search for localized excesses}\label{sec:locXS}
The first analysis is a blind search for excesses over the fraction of the sky covered with the Observatory. The number of UHECRs detected in circular windows on the sky ($N_\text{obs}$) is compared to that expected, in the same window, from an isotropic distribution of events ($N_\text{exp}$).  This search is performed over the entire field of view, which covers about 85\% of the sky. The search windows are centered on a \texttt{HEALPix} grid \citep[\texttt{HEALPix v3.70},][]{2005ApJ...622..759G}, defined by the parameter $\texttt{nSide} = 64$, which sets the size of the pixels to be of the order of the angular resolution of the Observatory. Events are counted within search windows of radius $\Psi$, ranging from $1^\circ$ to $30^\circ$ in $1^\circ$ steps. Similarly, the search is performed by selecting events above energy thresholds, $E_{\rm th}$, ranging from $32\:$EeV to $80\:$EeV in $1\:$EeV steps.
For each window and energy threshold, we estimate the binomial probability of obtaining by chance $N_\text{obs}$ or more events from an isotropic distribution of data. The computation of $N_\text{exp}$ is performed by simulating events with coordinates distributed according to the sum of the vertical and inclined exposures, weighted in proportion to the observed number of events at energies above 32\:EeV (see Section~\ref{sec:data}). For each realization of the simulated data set, the number of events is of the same size as observed across the field of view. Simulated events follow the same energy distribution as the observed events. Performing the analysis on simulated isotropic data sets allows us to take into account the trial factors for having tested different directions, radii and energy thresholds. We consider as \textit{post-trial} probability the fraction of these simulations with an equal or lower local \textit{p}-value than the best one obtained with the observed data set.

\begin{figure}[t]
\epsscale{1.2}
\plotone{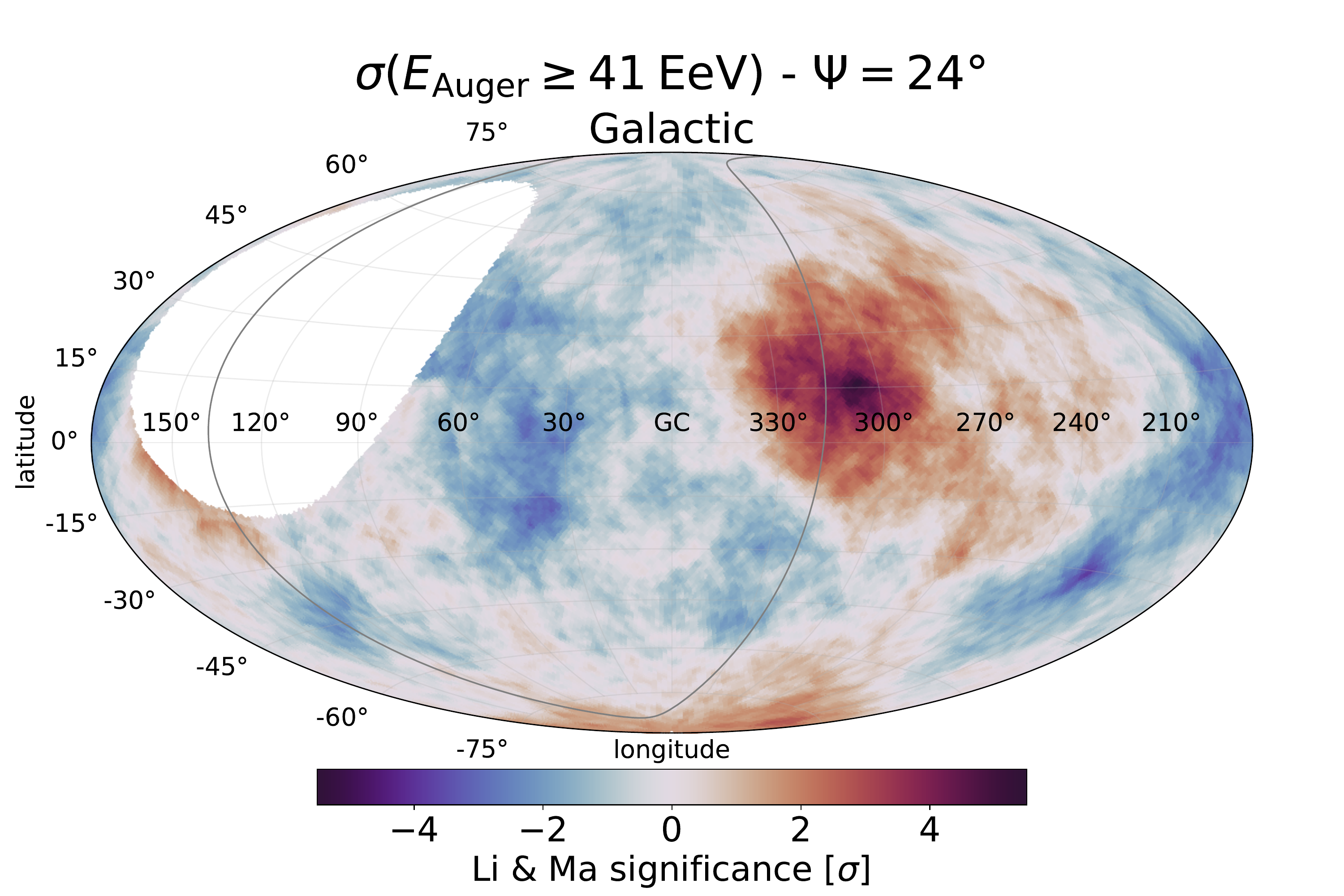}
\caption{Local Li-Ma significance map at energies above 41\:EeV and within a top-hat search angle $\Psi = 24^\circ$ in Galactic coordinates. The supergalactic plane is shown as a gray line. The significance is not evaluated in windows whose centers lie outside of the field-of-view of the Observatory, as indicated by the white area. \label{fig:blind}}
\end{figure}

We also compute the local Li-Ma significance \cite[equation~(17) in][]{1983ApJ...272..317L} for each point in the sky, where the ON-region is centered on each point of the \texttt{HEALPix} grid and the OFF-region is defined as the remainder of the field of view. The local significance map is displayed in Galactic coordinates in Figure~\ref{fig:blind}. The most significant excess, with $5.4\: \sigma$ local significance, is found above an energy threshold of $41\:$EeV within a top-hat window of $24^\circ$ radius centered on equatorial coordinates $(\alpha, \delta)=(196.3^\circ,-46.6^\circ)$, which corresponds to Galactic coordinates  $(l, b)=(305.4^\circ,16.2^\circ)$. At this position of the parameter space, 153 events are observed when 97.7 are expected from isotropy. The local $p$-value in this position is $3.7 \times 10^{-8}$, resulting in a post-trial $p$-value of 3\%. 

\subsection{Autocorrelation} \label{sec:autoc}
Another model-independent approach to assess the clustering of events is the search for autocorrelation, i.e.\ counting pairs of events separated by a given angular distance. This approach is particularly effective if the events form multiple clusters on similar angular scales in different directions in the sky. 

\begin{figure*}[t]
\gridline{\fig{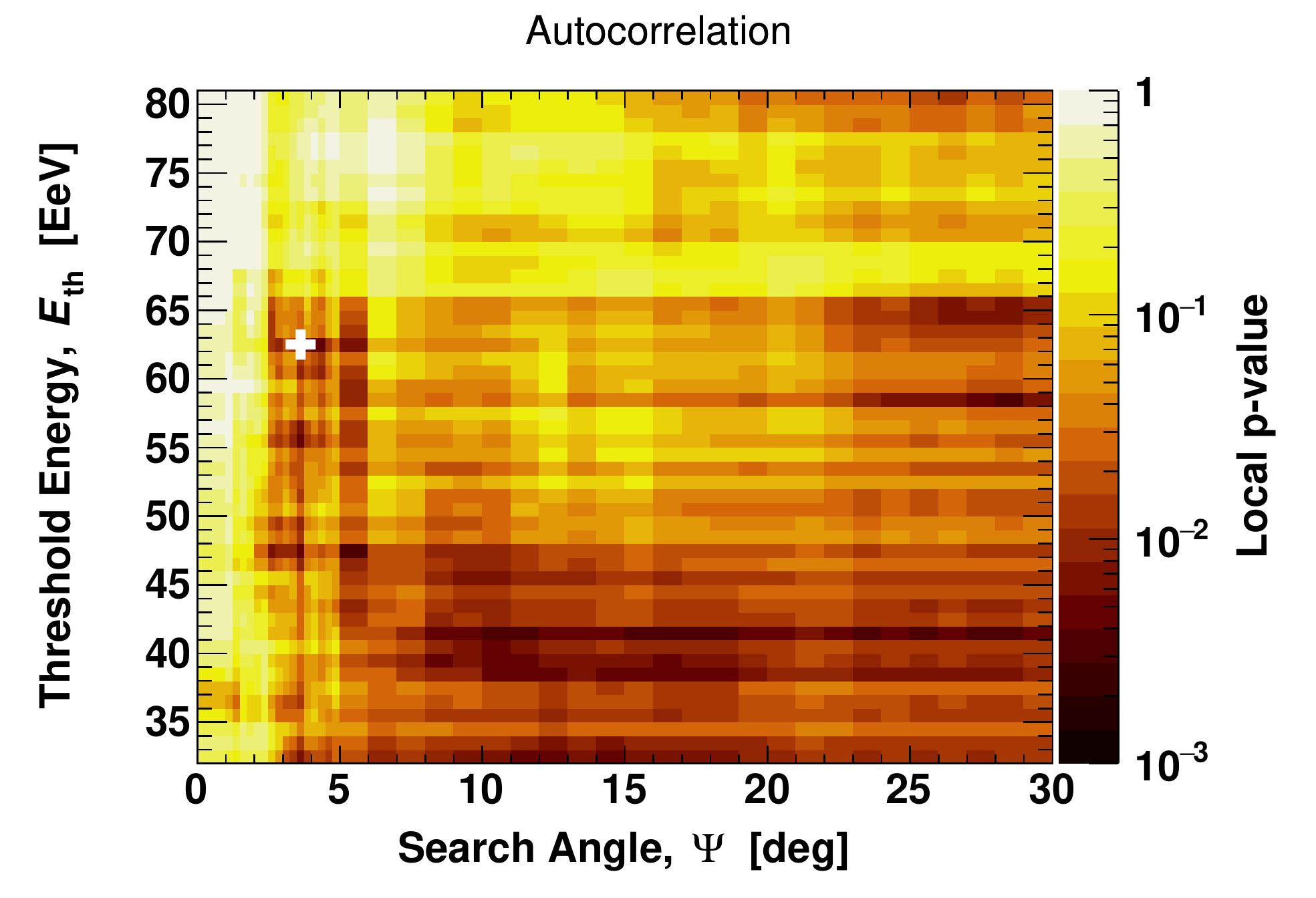}{0.46\textwidth}{(a)}
          \fig{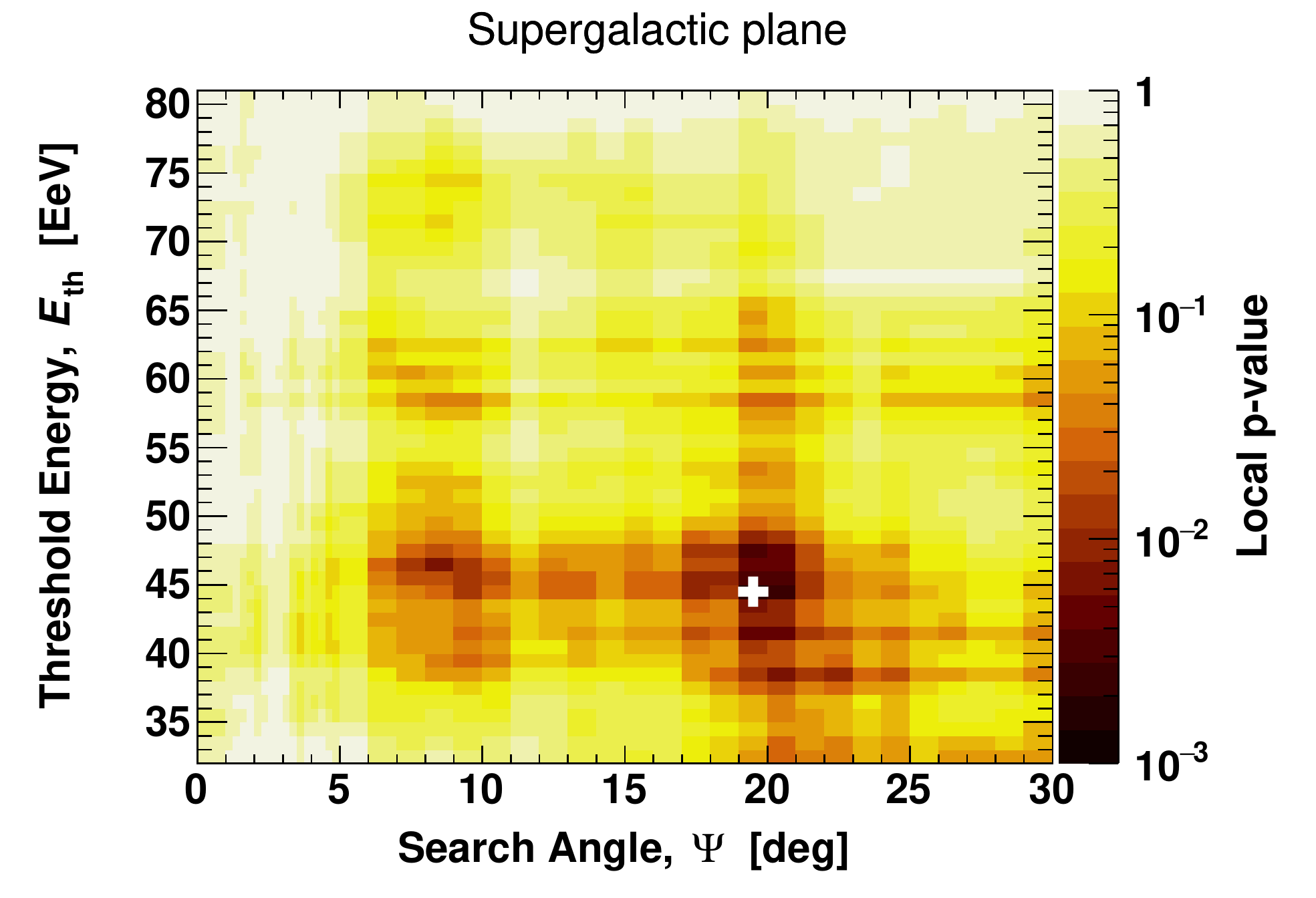}{0.46\textwidth}{(b)}
}
\vspace{-0.3cm}
\gridline{\fig{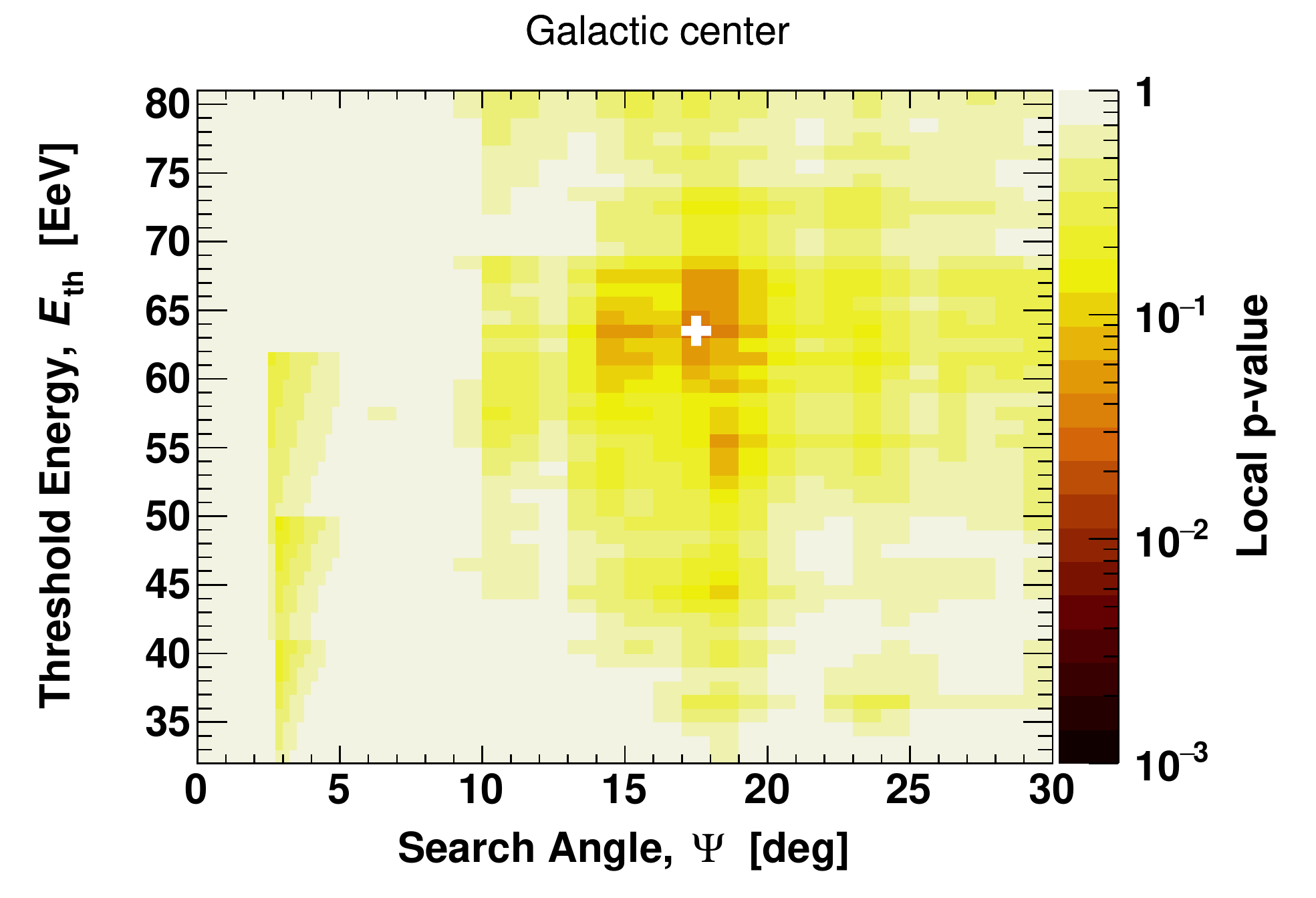}{0.46\textwidth}{(c)}
          \fig{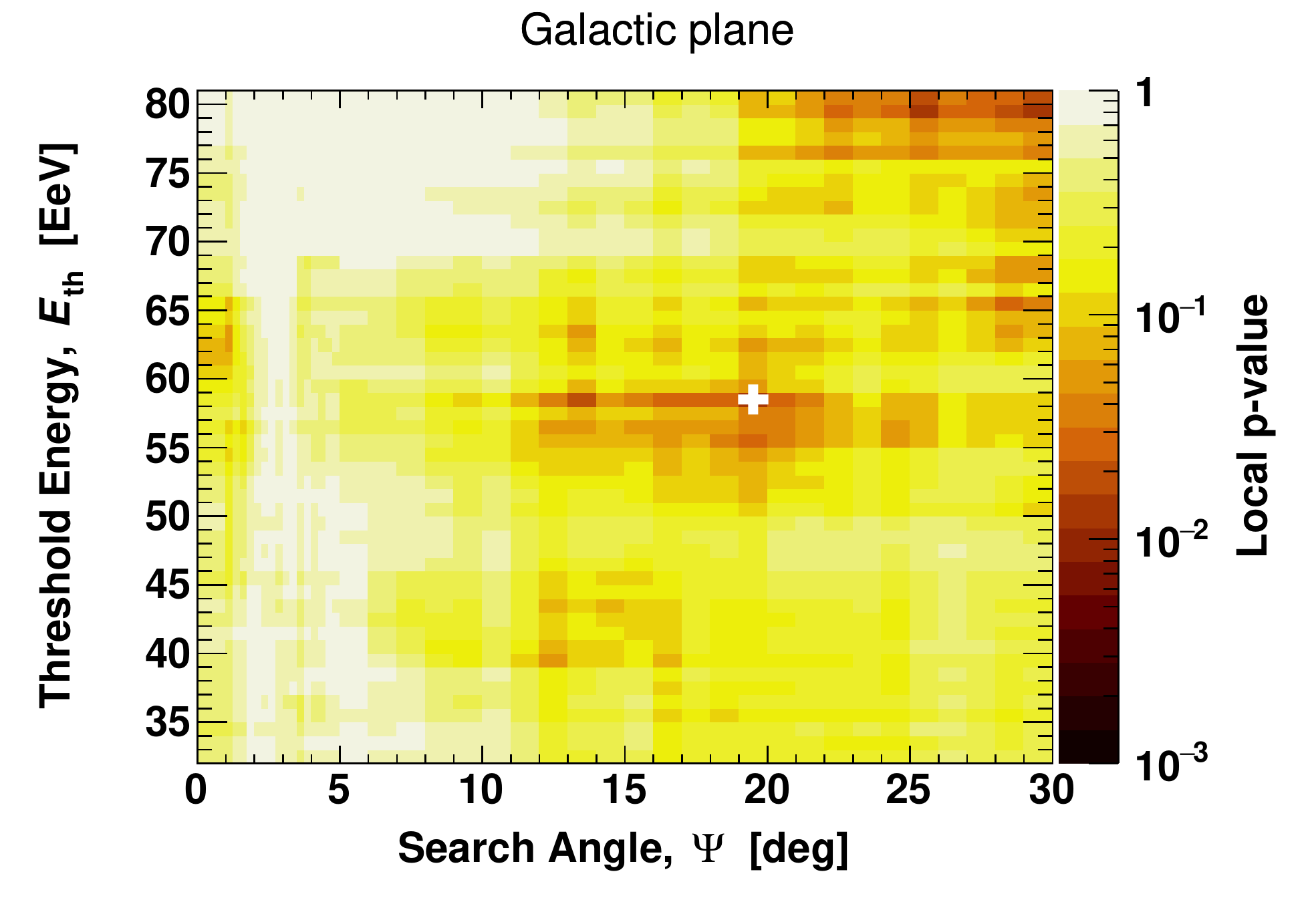}{0.46\textwidth}{(d)}
          }
\caption{Local $p$-value as a function of search angle, $\Psi$, and threshold energy, $E_\text{th}$. Panels a, b, c, d display the results of the autocorrelation study, supergalactic-plane, Galactic-center and Galactic-plane searches, respectively. The most significant excess identified in each analysis is indicated with a white cross. \label{fig:structures}}
\end{figure*}

Following \cite{2015ApJ...804...15A}, we count the number of event pairs, $N_\text{obs}$, above energy thresholds ranging from 32 to $80\:$EeV, that are separated by less than an angle $\Psi$ ranging from $1^\circ$ to $30^\circ$ in steps of $0.25^\circ$ up to $5^\circ$ and of $1^\circ$ above. We compute the expected number of pairs, $N_\text{exp}$, by analysing simulated isotropic event sets of the same size as the observed data set. For each $\Psi$ and $E_\text{th}$, we consider as local $p$-value the fraction of simulated data sets, $f(E_\text{th},\Psi)$, for which $N_\text{exp} \geq N_\text{obs} $. The values of $f$ are shown in Figure~\ref{fig:structures}(a) and the best results are shown in Table~\ref{tab:structures}. 

\newpage
\subsection{Correlation with structures} \label{sec:struc}
The most constrained analysis performed in this Section is a search for correlation with local astrophysical structures. Although a Galactic origin of UHECRs at energies above 8\:EeV is disfavored by the large-scale anisotropy discovered by the Collaboration, we test as targets the Galactic plane and the Galactic center in addition to the supergalactic plane, for consistency with \cite{2015ApJ...804...15A}. The search is performed in a similar way as the study described in Section~\ref{sec:autoc}, with $N_\text{obs}$ being the number of events observed within an angle $\Psi$ from the chosen structure. In practice, for the Galactic and supergalactic planes, we count events with an absolute value of latitude smaller than $\Psi$ in the respective coordinate system. 

The results are shown in Figure~\ref{fig:structures} and in Table \ref{tab:structures}. The lowest $p$-values are found for $\Psi \lesssim 20^\circ$ above energy thresholds near ${\sim}\:40$ and ${\sim}\:60$\:EeV. No significant departure from isotropy is observed in these searches, as in \cite{2015ApJ...804...15A}.

\begin{deluxetable*}{ccccccc}[ht]
\tablecaption{The results of the search for autocorrelation and correlation with astrophysical structures. \label{tab:structures}}
\tablehead{
\colhead{Search}	 	&	\colhead{$E_\text{th}$ [EeV]} &	\colhead{Angle, $\Psi$ [deg]} 	&	\colhead{$N_\text{obs}$} 	&	\colhead{$N_\text{exp}$} 	&	\colhead{Local $p$-value, $f_{\rm min}$} 	&	\colhead{Post-trial $p$-value}}
\startdata
Autocorrelation 	    &	62	&	3.75	&	$\phantom{000,}93$	        &	$\phantom{000,}66.4$	    &	 $2.5 \times 10^{-3}$ 	&	0.24	\\
Supergalactic plane 	&	44	&	$\phantom{0,}$20	&	$\phantom{00,}394$	&	$\phantom{00,}349.1$	&	 $1.8 \times 10^{-3}$  &	0.13	\\
Galactic plane 	        &	58	&	$\phantom{0,}$20	&	$\phantom{00,}151$	&	$\phantom{00,}129.8$	&	 $1.4 \times 10^{-2}$	&	0.44	\\
Galactic center	        &	63	&	$\phantom{0,}$18	&	$\phantom{000,}17$	&	$\phantom{000,}10.1$	&	 $2.6 \times 10 ^{-2}$	&	0.57	\\
\enddata
\tablecomments{The energy threshold, $E_{\rm th}$, and the search angle, $\Psi$, which minimize the local $p$-value, based on the number of observed and expected events / pairs. The post-trial $p$-value accounts for the scan in energy threshold and search angle, $\Psi$.}
\end{deluxetable*}

\section{Likelihood analysis with catalogs of candidate host galaxies } \label{sec:likelihood}

In \cite{2015ApJ...804...15A}, we presented the results of cross-correlation studies with three flux-limited catalogs: the 2MASS Redshift Survey of near-infrared galaxies \citep{2012ApJS..199...26H}, the \textit{Swift}-BAT 70-month catalog of active galactic nuclei (AGNs) observed in hard X-rays \citep{2013ApJS..207...19B} and a catalog of radio-emitting galaxies from \cite{2012A&A...544A..18V}. Such cross-correlation analyses inherently assume all galaxies under investigation to have an equal weight (standard-candle approach) and do not easily account for the inverse-square law of the UHECR flux, nor for its attenuation resulting from energy losses induced by propagation. These limitations were addressed in \cite{2018ApJ...853L..29A} through a likelihood-ratio test that expanded upon the maximum-likelihood test presented in \cite{2010APh....34..314A}. We also tested two additional catalogs based on gamma-ray observations from \textit{Fermi}-LAT. The full-sky gamma-ray survey of \textit{Fermi}-LAT has shown starforming galaxies and jetted AGN to be the main contributors to the extragalactic gamma-ray background at GeV energies, although their relative contributions remains uncertain  \cite[see e.g.][]{2015ApJ...800L..27A,2021Natur.597..341R}.

\subsection{From catalogs to UHECR sky models}
\label{sec:cat}

We first explore correlations with the large-scale distribution of matter using the Two Micron All-Sky Survey \citep[2MASS, ][]{2006AJ....131.1163S}. The expected UHECR flux in this scenario is traced by K-band observations at 2.16\:\textmu{m}, i.e.\ we assume an UHECR luminosity proportional to stellar mass. We limit the study to galaxies up to a K-band magnitude of 11.75\:mag, which corresponds to the flux limit over more than 90\% of the 2MASS Redshift Survey. We verified through the HyperLEDA\footnote{\url{http://leda.univ-lyon1.fr/}} database \citep{2014A&A...570A..13M} that all the selected objects are galaxies and we kept in the sample AGN hosts, noting though that their near-infrared emission may be contaminated by non-thermal emission.

A second sample consists of galaxies with a high star-formation rate, broadly denoted here as starburst galaxies. \cite{2019JCAP...10..073L} selected local galaxies with a far-infrared flux at 60\:\textmu{m} larger than 60\:Jy from the IRAS all-sky survey \citep{2003AJ....126.1607S} and with a radio flux at 1.4\:GHz larger than 20\:mJy from the NVSS \citep{1998AJ....115.1693C} and Parkes surveys \citep{2014PASA...31....7C} in the Northern and Southern hemispheres, respectively. The authors also imposed a far-infrared to radio flux ratio larger than 30, which removes galaxies dominated by jetted AGN emission. We further select galaxies with a far-infrared to radio flux ratio smaller than 1000, which excludes dwarf galaxies with negligible radio emission. The latter criterion removes the Large and Small Magellanic Clouds from the sample of starburst galaxies in \cite{2019JCAP...10..073L}, as these are clear outliers of the flux-ratio distribution. Although the IRAS survey can safely be considered as flux limited over the entire sky for fluxes larger than 60\:Jy, the subtraction of the Galactic foreground is more demanding in studies of extended radio sources down to 20\:mJy. Following their reanalysis of the Southern radio sky, \cite{2019JCAP...10..073L} excluded areas close the Galactic plane, which contain in particular the bright Circinus galaxy at latitude $l = -3.8^\circ$. The latter galaxy satisfies the above-mentioned selection criteria and we add it to the sample using its radio flux tabulated in the Parkes catalog \citep{1996yCat.8015....0W}. The radio flux of galaxies in the sample is used as a tracer for UHECR emission, effectively assuming an UHECR luminosity proportional to starforming activity.

The third sample encompasses AGNs observed in hard X-rays with \textit{Swift}-BAT, as tabulated in their 105-month catalog \citep{2018ApJS..235....4O}. We select hard X-ray sources with a $14-195\:$keV flux larger than $8.4\times10^{-12}$\:erg\:cm$^{-2}$\:s$^{-1}$, which corresponds to the \textit{Swift}-BAT flux limit over more than 90\% of the sky. We retain objects labeled as jetted AGN, Seyfert galaxies, or other AGNs with or without jets. We adopt with this catalog the hard X-ray flux as a tracer for the UHECR flux, effectively assuming that the UHECR luminosity is driven by accretion onto super-massive black holes. We note though that the X-ray flux of the sub-sample of radio-loud AGN, in particular that of blazars, is expected to be dominated by jet emission.

Finally, the fourth sample comprises $\gamma$-ray selected AGN from the \textit{Fermi}-LAT 3FHL catalog \citep{2017ApJS..232...18A}. We select radio galaxies and blazars with an integral flux between 10\:GeV and 1\:TeV larger than $3.3\times10^{-11}\:$cm$^{-2}$\:s$^{-1}$. Above this value, the 3FHL catalog is flux-limited over 90\% of the sky (97\% for Galactic latitudes $|b| > 5^\circ$).\footnote{Estimated from the data in Figure~6 of \cite{2017ApJS..232...18A}, where the flux limit is provided for a source of photon index $\Gamma=2.5$ detected with a test statistic $\text{TS}=25$. Data in the figure courtesy of the \textit{Fermi}-LAT Collaboration.} The $\gamma$-ray flux is used as UHECR proxy, effectively assuming an UHECR luminosity proportional to the inner jet activity.

The bands adopted to trace UHECR emission are affected by little absorption in the host galaxy and along the line of sight but UHECRs suffer increasing energy losses and photo-dissociation with increasing travel time. Robust estimates of the luminosity  distances of host galaxies are needed to account for the attenuation of their relative UHECR flux above a given energy threshold. Putative sources within a few tens of Mpc may in particular have a substantial impact on UHECR anisotropies while their host galaxies are not in the Hubble flow, which would make their spectroscopic redshift a biased distance estimate. We cross-matched all four catalogs with the HyperLEDA database and adopted the best distance estimate (\texttt{modbest} field) and associated uncertainty, which account for peculiar motion and exploit cosmic-distance-ladder estimates whenever available. Galaxies within 250\:Mpc are retained in the sample and we exclude those located in the Local Group through a cut at 1\:Mpc. Nearby galaxies would otherwise dominate sky models aimed at tracing UHECR emission on larger scales. A smaller horizon at 130\:Mpc is considered for starburst galaxies, following the selection of \cite{2019JCAP...10..073L}. We note  that few (if any) starforming galaxies within $130-250$\:Mpc are expected to pass the radio and far-infrared flux selection. All 26 jetted AGNs and 44 starburst galaxies in our sample are included in HyperLEDA. The apparent total K-band magnitude available in HyperLEDA  (\texttt{Kt} field) enables a straightforward selection of 44,113 2MASS galaxies. We identified 23 \textit{Swift}-BAT AGN, among 523 host galaxies, without a tabulated HyperLEDA distance that nonetheless show compatible redshift estimates ($|\Delta z| < 0.002$) in NED\footnote{doi: \href{https://ned.ipac.caltech.edu/}{10.26132/NED1}} and SIMBAD.\footnote{\url{http://simbad.u-strasbg.fr}} The distances of these 23 galaxies are based on their NED spectroscopic redshifts (corrected for the Local-Group infall to the Virgo cluster), as tabulated in Appendix~\ref{app:catalogs}. 

As in \cite{2018ApJ...853L..29A}, the UHECR flux expected from each host galaxy is increasingly attenuated with increasing luminosity distance, $d_\text{L}$, following the best-fit model of the spectrum and composition data acquired at the Pierre Auger Observatory \citep[][first minimum obtained with the EPOS-LHC hadronic interaction model]{2017JCAP...04..038A}. The attenuation weights, $a(d_\text{L})$, are marginalized over distance uncertainty for the three catalogs with less than 1,000 galaxies, with little impact on the final sky models. For the sake of computational intensity, no marginalization over distance uncertainty is performed for the fourth sample, made of more than 44,000 near-infrared galaxies, with negligible impact on the final results.

All four sky models represent significant improvements with respect to those studied in \cite{2018ApJ...853L..29A} from an astronomical point of view. From a quantitative perspective, the improvement in sky coverage and depth of the surveys yield an increase in jetted AGN from 17 to 26 objects, in starburst galaxies from 23 to 44, in all AGNs from 330 to 523 and in near-infrared galaxies from 41,129 to 44,113. The estimation of distance uncertainties also provides a qualitative improvement with respect to the study presented in \cite{2018ApJ...853L..29A}. It should be noted though that the results are barely affected by such improvements (see Section~\ref{sec:likelihood_res}), suggesting that our previous analysis already accounted for sufficiently complete surveys from an astroparticle point of view.

We further evaluated in \cite{2015ApJ...804...15A} possible correlations with the catalog of \cite{2012A&A...544A..18V}. The latter compiles observations at 1.4\:GHz and 843\:MHz of extended radio sources down to a flux limit corresponding to the flux of Centaurus A placed at 200\:Mpc. Accounting for attenuation, the resulting sky model is entirely dominated by the nearby Centaurus A (distance of $3.68\pm0.05\:$Mpc) and can thus be considered as redundant with the flux pattern obtained with the \textit{Swift}-BAT model (see Appendix~\ref{app:catalogs}). We thus limit the present study to the four sky models obtained from near-infrared emission of galaxies (2MASS), radio emission from starburst galaxies, X-rays from AGNs (\textit{Swift}-BAT) and $\gamma$-rays from jetted AGNs (\textit{Fermi}-LAT).

\subsection{Likelihood-ratio analysis}
\label{sec:maxL}

As in \cite{2018ApJ...853L..29A}, the correlation of UHECR arrival directions with the flux pattern expected from the catalogs is evaluated against isotropy using a likelihood-ratio analysis. The model as a function of direction ${\bf u}$ is computed in equal-area bins on the sphere using {\tt HEALPix v3.70} with the parameter {\tt nSide = 64}, as in Section~\ref{sec:locXS}.

The null hypothesis under investigation, $H_0$, is that of an isotropic flux distribution. Accounting for the directional exposure of the array, $\omega({\bf u})$, the isotropic model for the UHECR count density reads
\begin{equation}
n^{H_0}({\bf u}) = \frac{\omega({\bf u})}{\sum_i \omega({\bf u}_i)},
\end{equation}
which is normalized so that the sum over the {\tt HEALPix} pixels indexed over $i$ and of direction ${\bf u}_i$ is equal to one.

The alternative hypothesis, $H_1$, in which $H_0$ is nested, is considered as the sum of an isotropic component and a component derived from the tested catalog. The amplitude of the latter component is a variable signal fraction, $\alpha$. The isotropic remainder accounts for faint or distant galaxies not included in  the catalogs or for a heavy nuclear component deflected away on large angular scales. The model for the UHECR count density under $H_1$ reads
\begin{equation}
\label{eq:likelihood_model}
n^{H_1}({\bf u}) = (1-\alpha)\times n^{H_0}({\bf u}) + \alpha\times\frac{\sum_j s_j({\bf u}; \Theta)}{\sum_i \sum_j s_j({\bf u_i}; \Theta)},
\end{equation}
where the index $j$ runs over the galaxies in the catalog. The contribution to the UHECR flux from each galaxy, $s_j({\bf u}; \Theta)$, is modeled as a von Mises-Fisher distribution centered on the direction of the galaxy with a smearing angle $\Theta$. The amplitude of its contribution is proportional to the electromagnetic flux of the galaxy, $\phi_j$, accounting for attenuation as a function of luminosity distance, $a(d_j)$, so that
\begin{equation}
\label{eq:Fisher}
    s_j({\bf u}; \Theta) = \omega({\bf u})\times\phi_ja(d_j)\times\exp\Bigg(\frac{{\bf u}\cdot{\bf u}_j}{2(1-\cos \Theta)}\Bigg).
\end{equation}
The von Mises-Fisher distribution is maximum in the direction of the galaxy of interest, ${\bf u}_j$, effectively leaving aside coherent deflections which remain under-constrained by current models of the Galactic magnetic fields \citep{2016APh....85...54E}. The smearing angle $\Theta$, equivalent to the 2D Gaussian extent in the small-angle limit, is assumed to be the same for all galaxies in a given catalog. This parameter accounts for the average angular dispersion in intervening magnetic fields. As a note, the normalization of the von Mises-Fisher distribution in equation~\eqref{eq:Fisher} is omitted, as it is the same for every galaxy and as the overall anisotropic component is normalized on the sphere (see equation~\eqref{eq:likelihood_model}).

The likelihood-ratio test between the nested models $H_0$ and $H_1$ defines the test statistic, ${\rm TS} = 2\ln ( {\cal L}_1/{\cal L}_0 )$, where the likelihood scores of the null and alternative hypothesis, ${\cal L}_0$ and ${\cal L}_1$, are  obtained as the product over the events of the models $n^{H_0}$ and $n^{H_1}$, respectively. The evaluation of the test statistic is performed by grouping events by {\tt HEALPix} bin. With an observed event count $k_i$ in the direction  ${\bf u}_i$, the test statistic is evaluated as
\begin{equation}
{\rm TS} = 2\sum_i k_i \times \ln\frac{n^{H_1}({\bf u}_i)}{n^{H_0}({\bf u}_i)}.
\end{equation}

The test statistic is maximized as a function of the two free parameters in the analysis (the search radius, $\Theta$, and the signal fraction, $\alpha$) above successive energy thresholds. The maximization can be achieved by scanning the 2D parameter space by steps of 0.2\% in signal fraction and $0.2^\circ$ in search radius. This approach provides an accurate estimate that is independent from any specific maximization algorithm. Alternatively, a maximization with the {\tt Minuit} package provides a fast estimate for simulated data sets, with an accuracy on TS better than 0.1 units for event counts larger than 100. Above a fixed energy threshold, the test statistic is observed through Monte Carlo simulations to follow a $\chi^2$ distribution with two degrees of freedom under the null hypothesis \citep{wilks1938}. The 1 and $2\: \sigma$ C.L.~on the best-fit parameters are set by iso-TS contours differing from the maximum TS value by 2.3 and 6.2 units, respectively.

The scan in energy threshold is accounted for, as in Section~\ref{sec:blind}, by estimating the post-trial $p$-value through isotropic Monte Carlo simulations. The post-trial $p$-value, which accounts for the energy scan, differs from the local $p$-value expected from Wilks' theorem by a penalty factor that is well-approximated by a linear function of TS: ${\rm pen} =1+(0.30\pm0.01)\times{\rm TS}$. This empirical penalty factor is estimated from simulated isotropic data sets analyzed against each catalog and the uncertainty on the linear coefficient is estimated from the variance across the four tested catalogs. The penalty factor reaches a value of ${\sim}\:10$ for ${\rm TS} = 30$. 

\subsection{Results}
\label{sec:likelihood_res}

\begin{figure}
\epsscale{1.22}
\plotone{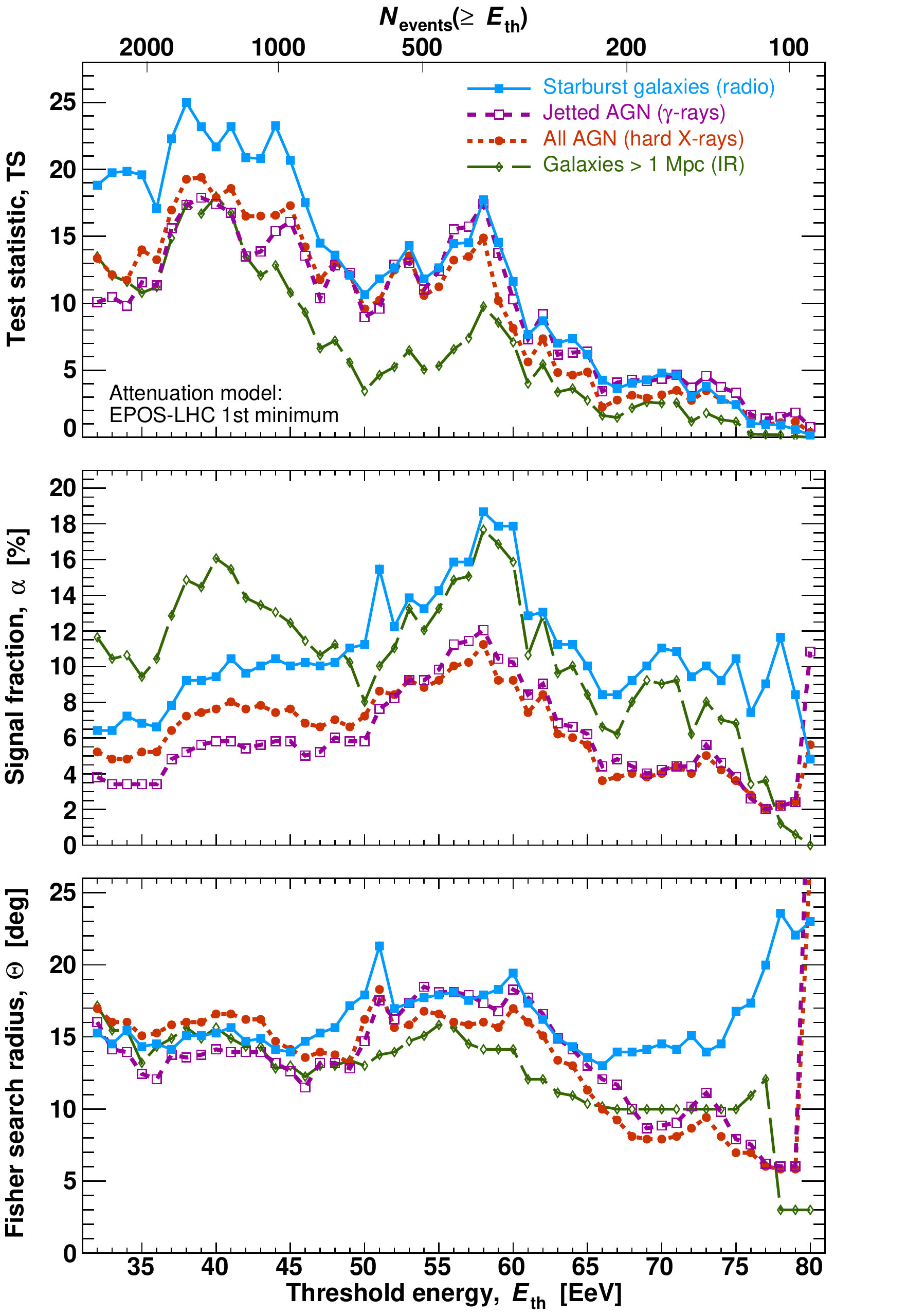}
\caption{The test statistic (top), signal fraction (center) and Fisher search radius (bottom) maximizing the deviation from isotropy as a function of energy threshold. The results obtained with each of the four catalogs are displayed with varying colors and line styles, as labeled in the Figure. The uncertainties on the parameters, which are correlated above successive energy thresholds, are not displayed for the sake of readability.\label{fig:CatRes}}
\end{figure}

The search radius and signal fraction maximizing the test statistic above fixed energy thresholds ranging in 32--80\:EeV are displayed in Figure~\ref{fig:CatRes} for the four catalogs. The test statistic follows a double hump structure as a function of energy, with a first peak at energies above ${\sim}\:40$\:EeV and a second peak at energies above ${\sim}\:60$\:EeV. The latter peak corresponds to the maximum signal fraction for all catalogs, ranging in 11--19\%. Lower signal fractions ranging in 6--16\% are inferred from the global TS maximum, at energies above ${\sim}\:40$\:EeV. As shown in the upper axis in Figure~\ref{fig:CatRes}, the four times larger number of events in the first peak  (1,387 above 40\:EeV vs 331 above 60\:EeV) yields a more significant deviation from isotropy above $40\:\text{EeV}$. 

\begin{figure*}[ht]
\epsscale{1.1}
\plotone{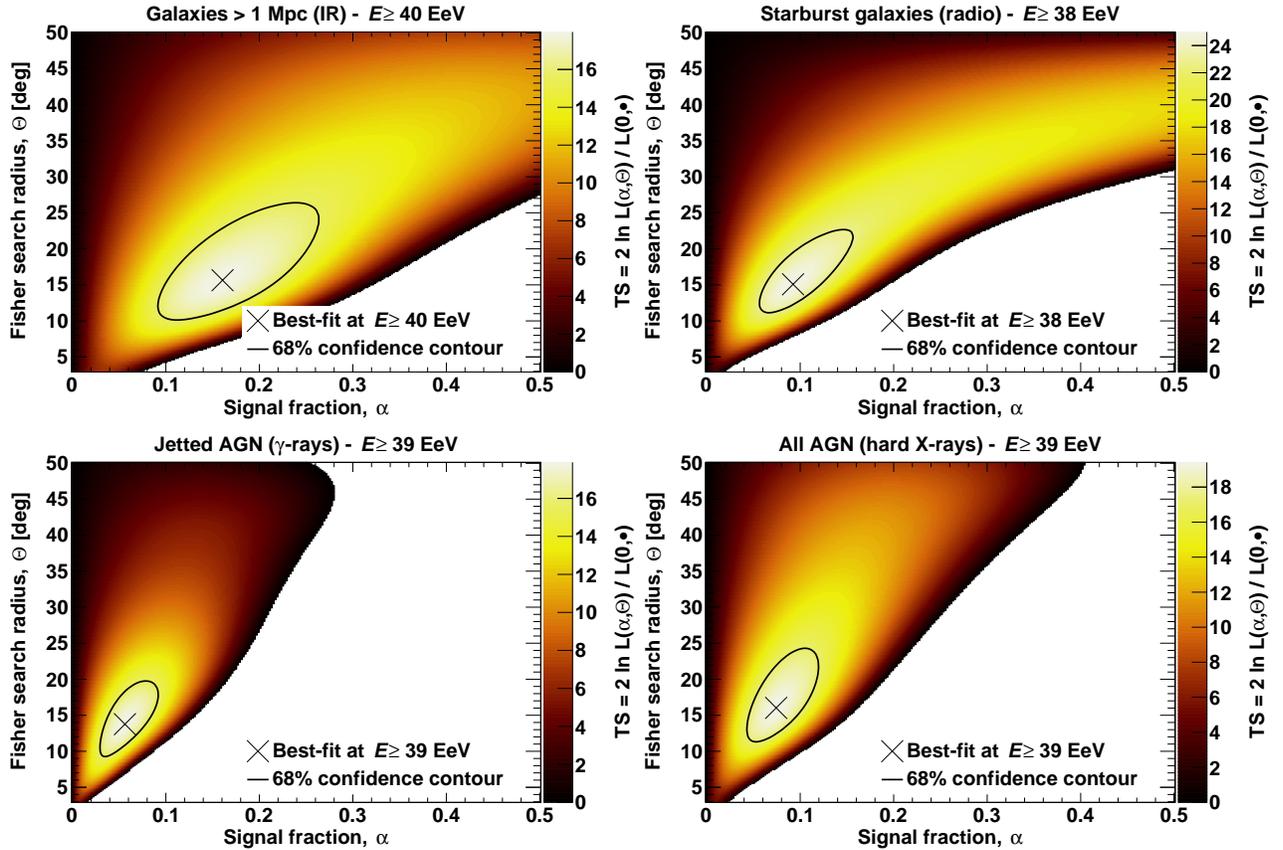}
\caption{The test statistic as a function of signal fraction and search radius for the four tested catalogs, as labeled in the Figure. The reference best-fit parameters obtained above the energy threshold that maximizes the departure from isotropy are marked with a cross. The 68\% C.L.~contour is displayed as a black line. The complete Figure set (4 $\times$ 49 images), which shows the evolution of the test statistic mapping as a function of energy threshold, is available in the online journal, in the arXiv source file and on the \augerwebsite\ of the Pierre Auger Collaboration.
\label{fig:bestfit_params}
}
\end{figure*}

The amplitude of variations of the best-fit parameters as function of energy threshold can be evaluated against the statistical uncertainties on these parameters, as shown in Figure~\ref{fig:bestfit_params}. As the search is performed above successive energy thresholds by steps of 1\:EeV,  successive energy bins have a non-negligible overlap. For reference, we estimate that there is a total of five to six independent energy bins, by identifying the successive reference energy thresholds above which the number of events is less that half that above a previous reference energy. Such a procedure suggests reference energy thresholds at $E \gtrsim 32, 40, 50, 60, 70, 80\:$EeV, with boundaries distant by more than $\Delta \log_{10} E = 0.06$, that corresponds to the energy resolution of $\pm 7\%$ relevant in the range covered here \citep{2020PhRvD.102f2005A}. As illustrated by the set of Figures above energy thresholds ranging in 32--80\:EeV (see online material attached to Figure~\ref{fig:bestfit_params}), the reconstructed parameters do not show  significant variations with energy.

For the sake of completeness, we provide the best-fit parameters and maximum test statistic obtained above energy thresholds corresponding to the global maximum at $E\gtrsim 40\:$EeV, in the upper part of Table~\ref{tab:1st_best}, as well as those obtained above the secondary maximum identified at $E\gtrsim 60\:$EeV, in the lower part of the same table. The most significant departure from isotropy is identified for all four catalogs at energy thresholds in the range  38--40\:EeV, with post-trial $p$-values of $8.3 \times 10^{-4}$, $7.9 \times 10^{-4}$, $4.2 \times 10^{-4}$ and $3.2 \times 10^{-5}$ for jetted AGNs traced by their $\gamma$-ray emission, galaxies traced by their near-infrared emission, all AGNs traced by their X-ray emission and starburst galaxies traced by their radio emission, respectively. As in \cite{2018ApJ...853L..29A}, we do not penalize for the test of the four catalogs, which all provide similar UHECR flux patterns. As a note, the infrared sample of galaxies contains a large fraction (more than 75\%) of each of the three other catalogs and only jetted AGN and starburst catalogs can be considered as strictly distinct galaxy samples.

\begin{deluxetable*}{cccccc}
\tablecaption{The best-fit results obtained with the four catalogs at the global (upper) and secondary (lower) maximum.\label{tab:1st_best}}
\tablehead{\colhead{Catalog} & \colhead{$E_{\rm  th}$ [EeV]} & \colhead{Fisher search radius, $\Theta$ [deg]} & \colhead{Signal fraction, $\alpha$ [\%]} & \colhead{TS$_{\rm max}$} & \colhead{Post-trial $p$-value}}
\startdata
All galaxies (IR)           & 40 & $16_{-6}^{+11}$              & $16_{-7}^{+10}$                       & 18.0 & $7.9 \times 10^{-4}$ \\
Starbursts (radio)          & 38 & $15_{-4}^{+8\phantom{0}}$    & $\phantom{0}9_{-4}^{+6\phantom{0}}$   & 25.0 & $3.2 \times 10^{-5}$ \\
All AGNs (X-rays)           & 39 & $16_{-5}^{+8\phantom{0}}$    & $\phantom{0}7_{-3}^{+5\phantom{0}}$   & 19.4 & $4.2 \times 10^{-4}$ \\
Jetted AGNs ($\gamma$-rays) & 39 & $14_{-4}^{+6\phantom{0}}$    & $\phantom{0}6_{-3}^{+4\phantom{0}}$   & 17.9 & $8.3 \times 10^{-4}$ \\
\hline
All galaxies (IR)           & 58 & $14_{-5}^{+9}\phantom{0}$    & $18_{-10}^{+13}$                      & $\phantom{0}$9.8 & $2.9 \times 10^{-2}$ \\
Starbursts (radio)          & 58 & $18_{-6}^{+11}$              & $19_{-9}^{+20}$                       & 17.7 & $9.0 \times 10^{-4}$ \\
All AGNs (X-rays)           & 58 & $16_{-6}^{+8\phantom{0}}$    & $11_{-6}^{+7\phantom{0}}$             & 14.9 & $3.2 \times 10^{-3}$ \\
Jetted AGNs ($\gamma$-rays) & 58 & $17_{-5}^{+8\phantom{0}}$    & $12_{-6}^{+8\phantom{0}}$             & 17.4 & $1.0 \times 10^{-3}$ \\
\enddata
\tablecomments{The energy threshold, $E_{\rm th}$, Fisher search radius, $\Theta$, and signal fraction, $\alpha$, which maximize the test statistic, TS$_{\rm max}$, for each of the catalogs. The post-trial $p$-value accounts for the energy scan and search over $\alpha$ and $\Theta$.}
\end{deluxetable*}

As discussed in Sec.~\ref{sec:cat}, all four sky models tested here are based on improved versions of the catalogs used in \cite{2018ApJ...853L..29A}, although with a mild impact on the significance of the results and no noticeable change in the best-fit parameters. The maximum test statistic is obtained at the same point of the parameter space using the catalogs of infrared galaxies, starburst galaxies, and X-ray AGNs from \cite{2018ApJ...853L..29A}, with TS values of 16.0, 23.1 and 18.0, respectively, differing by less than 2 units from the results in Table~\ref{tab:1st_best}. The most important change is observed for the gamma-ray catalog of jetted AGNs: the maximum TS (13.5) is obtained above ${\sim}\:60\:$EeV with the earlier catalog version based on the 2FHL catalog ($E_\gamma > 50\:$GeV), while it is obtained above ${\sim}\:40\:$EeV with the current version based on the 3FHL catalog  ($E_\gamma > 10\:$GeV). The change can be understood from the lower energy threshold of the 3FHL catalog, which reduces the relative flux of blazars beyond 100\:Mpc (Mkn\:421, Mkn\:501) with respect to the flux of local radio galaxies (Cen\:A, NGC\:1275, M\:87).

\section{The Centaurus Region}
\label{sec:Cen}

A visual inspection of the sky models displayed in Appendix~\ref{app:catalogs} highlights the main similarity between the four catalogs, namely a hotspot expected in the Auger field of view in the direction of the group of galaxies composed of the radio galaxy Centaurus A, the Seyfert galaxy NGC\:4945 and the starburst galaxy M\:83. These three galaxies, at distances of about 4\:Mpc, constitute one of the pillars of the so-called Council of Giants  \citep{2014MNRAS.440..405M} surrounding the Milky Way and Andromeda galaxy. Inspection of the two AGN models, tracing accretion through X-ray emission and jet activity through $\gamma$-ray emission, does not suggest bright secondary hotspots in other sky regions at the highest energies ($E\gtrsim 60\:$EeV), as the attenuation of the UHECR flux dramatically reduces the contribution from more distant galaxies. On the other hand, both the infrared model of stellar mass and the radio model of enhanced starforming activity suggest hotspots in the directions of other members of the Council of Giants: the starburst galaxies NGC\:253 and M\:82, which are the only two starburst galaxies currently detected at TeV energies.\footnote{\url{http://tevcat2.uchicago.edu/}} While M\:82 lies in the blind region of the Pierre Auger Observatory, which can only be observed with Telescope Array \citep{2018ApJ...867L..27A}, the contribution from NGC\:253  is responsible for the larger departure from isotropy obtained with the starburst model with respect e.g.\ to the X-ray AGN model (see Appendix~\ref{app:catalogs}). The infrared model instead yields a smaller test statistic than both the X-ray AGN and starburst models. Within the infrared model, the region of the Virgo cluster (at $d\sim 20\:$Mpc) would be brighter than the Centaurus region, which is in tension with the UHECR observations. Following the same procedure as in \cite{2018ApJ...853L..29A}, we performed a quantitative comparison between the four models to determine whether one of them is favored by the data against the others. The infrared, X-ray and $\gamma$-ray models fit the data at $E \geq 38-40\:$EeV worse than the starburst model with C.L.~${\lesssim}\: 3\:\sigma$. No firm evidence for a catalog preference is identified.

The deviation from isotropy suggested with all four galaxy catalogs is driven by a hotspot in the Centaurus region. This region shows an enhanced flux in all four sky models, arising mainly from Centaurus~A for the two AGN models, NGC\:4945 for the starburst model and from both galaxies in the infrared model. The peak direction of the UHECR hotspot, as identified through the blind search described in Section~\ref{sec:locXS}, points $2.9^\circ$ away from the main contributor to the starburst model, NGC\:4945, and $5.1^\circ$ away from the main contributor to the AGN models, Centaurus~A.
 
\begin{figure}[t]
\epsscale{1.2}
\plotone{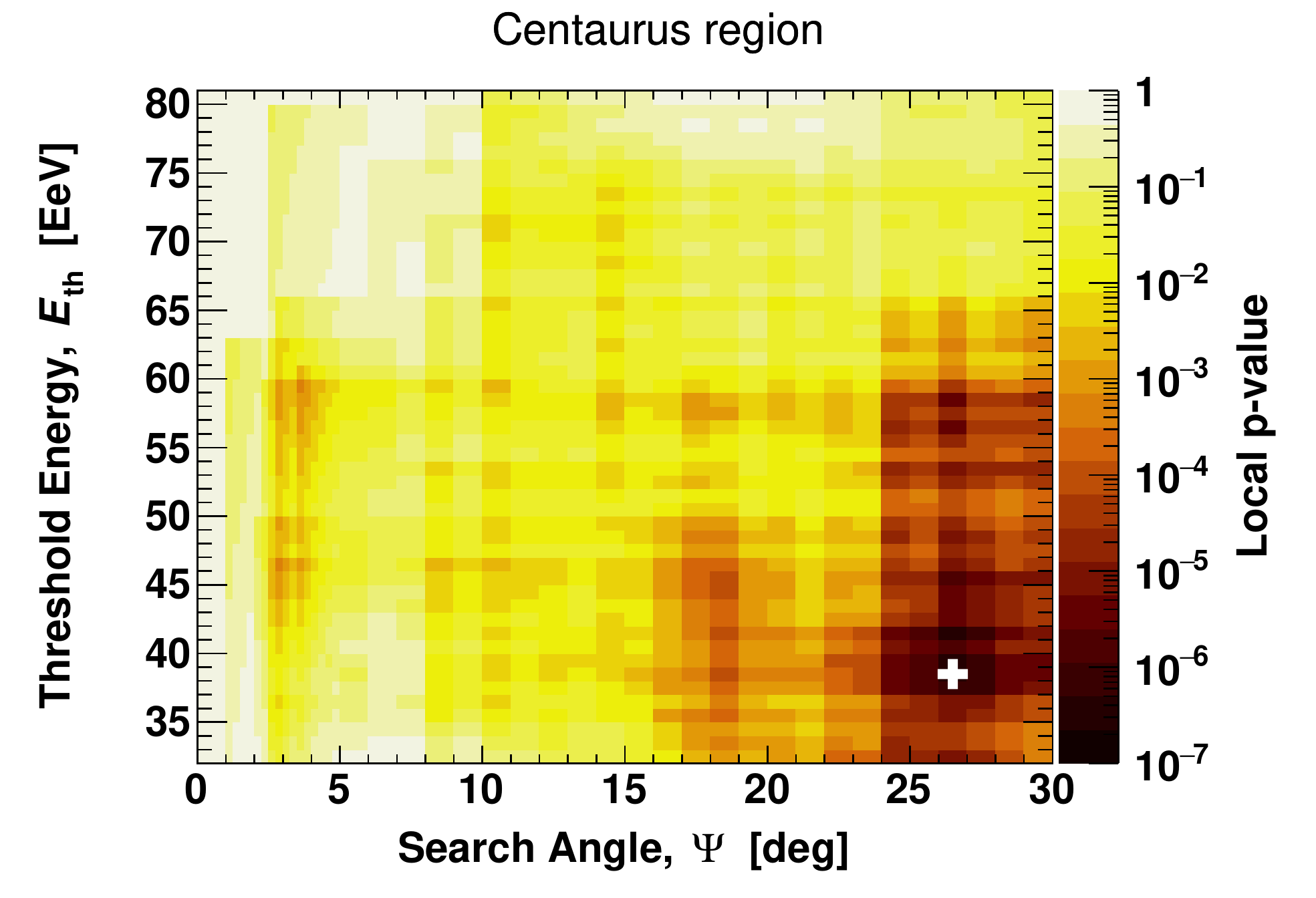}
\caption{The local $p$-value for an excess in the Centaurus region as a function of top-hat search angle and energy threshold. The minimum $p$-value, obtained for the best-fit parameters, is marked with a white cross. \label{fig:CenA}}
\end{figure}

Centaurus~A, being the closest radio galaxy at $3.68\pm0.05\:$Mpc, has been the target of searches for UHECR excess by the Pierre Auger Collaboration for more than a decade \citep{2007Science.1151124}. We update such searches by performing the same analysis described in Section~\ref{sec:struc} using as target the position of Centaurus~A, ($\alpha,\delta)=(201.4^\circ,-43.0^\circ)$. The map of the local $p$-values as a function of energy threshold and top-hat search angle is shown in Figure~\ref{fig:CenA}.  The most significant excess is found at $E_\text{th}=38\:$EeV in a circle of top-hat radius $\Psi=27^\circ$, where the number of observed events is $N_\text{obs}=215$ while $N_\text{exp}=152.0$ events would be expected from isotropy. The minimum local $p$-value, which is estimated as in Section~\ref{sec:blind} from the binomial probability to observe $N_\text{obs}$ or more events from an isotropic distribution, is $2.1 \times 10^{-7}$. After penalization for the scan in energy and search angle, the post-trial $p$-value is $4.5 \times 10^{-5}$, similar to that obtained with the likelihood-ratio test for starburst galaxies against isotropy. 

The best-fit parameters of the search in the direction of Centaurus~A are unsurprisingly similar to those of the blind search. The lower post-trial $p$-value with respect to the blind search results from the direction being fixed \textit{a priori}, as suggested by the early-day searches from the Pierre Auger Collaboration \citep{2007Science.1151124,2010APh....34..314A}.  The top-hat angular scale inferred from the blind search and from the search at the position of Centaurus~A, $\Psi=24-27^\circ$, can be compared to the Fisher search radius inferred from the catalog-based searches through the relation $\Psi = 1.59\times\Theta$.\footnote{For a  Fisher radius~$\Theta \ll 1$\:rad, this relation provides the top-hat radius~$\Psi$ that maximizes the signal-to-noise ratio, where the noise is~${\propto}\: \sqrt{1 - \cos\Psi}$ and the signal is~${\propto}\: \exp(k) -\exp(k\cos\Psi)$, with the concentration parameter~$k = \left[2(1 - \cos\Theta)\right]^{-1}$.} The catalog-based searches yield $\Theta = 14^\circ - 16^\circ$ that corresponds to $\Psi = 22^\circ - 25^\circ$, i.e.\ a range of values that is consistent with those inferred from the other searches. 

\begin{figure}[t]
\epsscale{1.2}
\plotone{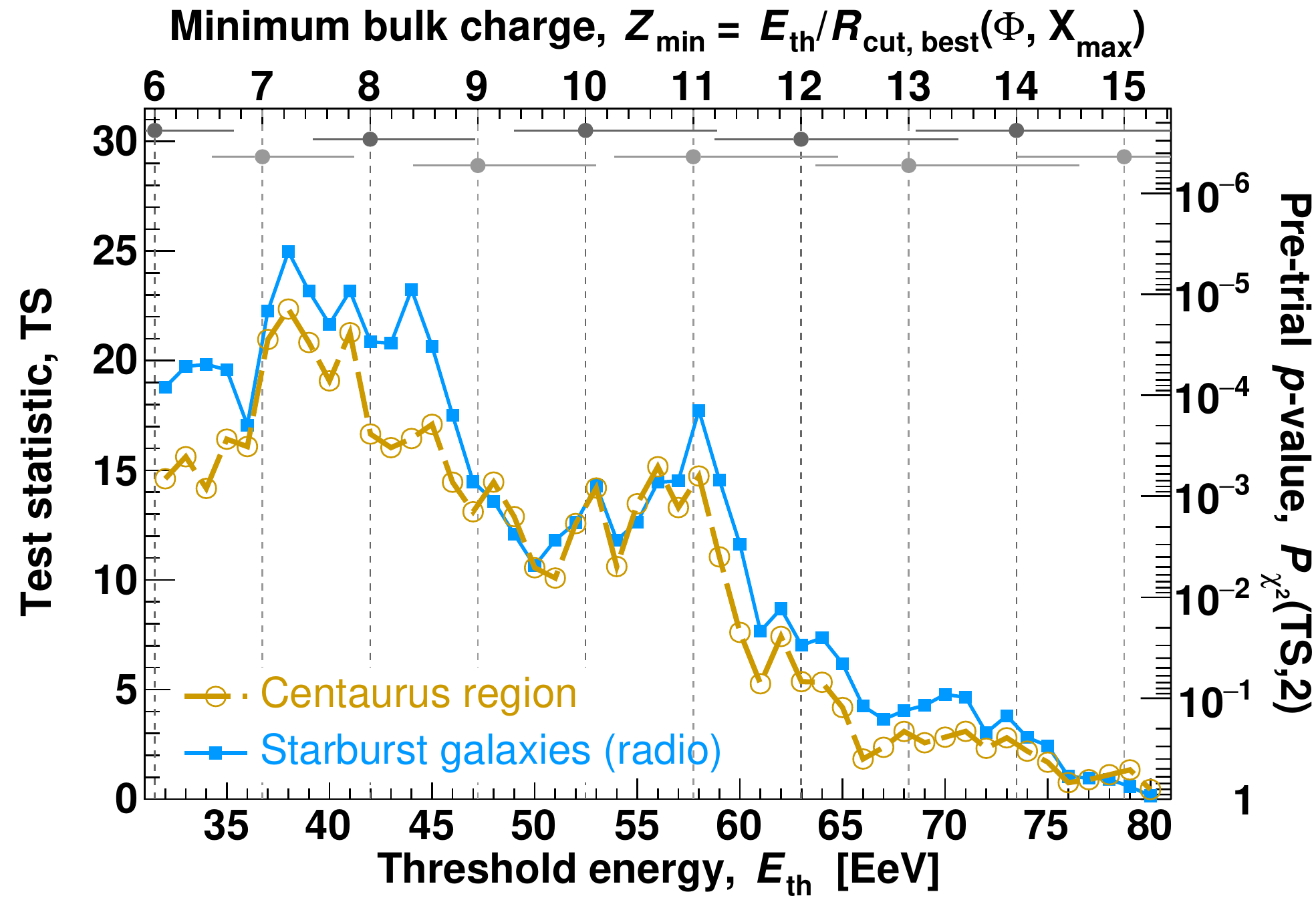}
\caption{The test statistic and pre-trial $p$-value, after profiling against the search radius and penalization for this free parameter, as a function of energy threshold. The gray points along the top axis figure the estimate of a lower bound on the bulk charge of UHECRs above a given energy threshold, under the assumption of an energy-to-charge ratio close to the maximum rigidity inferred by jointly modeling the energy spectrum and composition observables \citep{2017JCAP...04..038A}. \label{fig:Edependence}}
\end{figure}

Both the catalog-based searches and search in the Centaurus region point to a most significant signal at an energy threshold close to 40\:EeV. This energy range encompasses the flux suppression of the energy spectrum above the toe, at $E_{34} = 46 \pm 3 \pm 6\:$EeV \citep{2020PhRvD.102f2005A}. The evolution of the signal with energy displayed in Figure~\ref{fig:CatRes} appears to be mainly driven by the event distribution in the Centaurus region, as illustrated in Figure~\ref{fig:Edependence}. The pre-trial $p$-value in the Centaurus region is obtained by profiling the local $p$-value against the search radius and penalizing for this free parameter. The profile as a function of energy threshold is compared to the test statistic of the starburst catalog. The latter is chosen as example, noting that the results obtained with other catalogs show a similar dependence on energy threshold (see Figure~\ref{fig:CatRes}).

Constraints from maximum shower-depths up to a few tens of EeV and from the broad-band spectrum above the ankle energy suggest that UHECRs are accelerated in proportion to their charge, following so-called Peters' cycles \citep{2017JCAP...04..038A, PhysRevLett.125.121106}. The cosmic-ray composition above the toe in the energy spectrum is then expected to be dominated by UHECRs near a maximum magnetic rigidity, $R_{\rm cut}$. Accounting for both systematic uncertainties on the energy and maximum shower-depth scales, we inferred in \cite{2017JCAP...04..038A} a maximum rigidity $\log_{10}( R_{\rm cut} /\text{V}) = 18.72_{-0.03}^{+0.04}$ with our reference model. Adopting this value as the typical rigidity of UHECRs above the toe, a lower bound on the charge of the bulk of UHECRs above a given energy threshold can be estimated as $Z_{\rm min} = E_{\rm th}/R_{\rm cut}$, as figured in the top axis of Figure~\ref{fig:Edependence}. The uncertainties on the points illustrate those on the maximum rigidity in the reference scenario. It should be noted that the composition at the highest energies remains poorly constrained with Phase~1 data and can only be conjectured from a model-dependent approach at this stage. 

At rigidities close to $R_{\rm cut} = 5\:$EV, i.e.\ $\log_{10}( R_{\rm cut}/\text{V}) \approx 18.7$, UHECR propagation in the magnetic field of the Milky Way enters into a semi-ballistic regime \citep{2016APh....85...54E}. Excesses identified in the UHECR sky could thus be used both to track back putative sources and possibly to constrain the configuration and strength of the Galactic magnetic field \citep[see][and references therein]{2018JCAP...08..049B}. The angular scale inferred from the catalog-based search, as well as that from the blind search and search in the Centaurus region, are consistent with the average angular dispersion expected in the Milky Way of the Auger mix of nuclear species \citep{2018ApJ...853L..29A}. Nonetheless, the lack of a significant preference for a specific class of galaxies and the strength of the anisotropy signal, reaching at best post-trial $p$-values of $(3-5)\times10^{-5}$, still limit the identification of the host galaxies of UHECR accelerators and UHECR constraints on the Galactic magnetic field.

\begin{figure*}[ht!]
\epsscale{1.}
\plotone{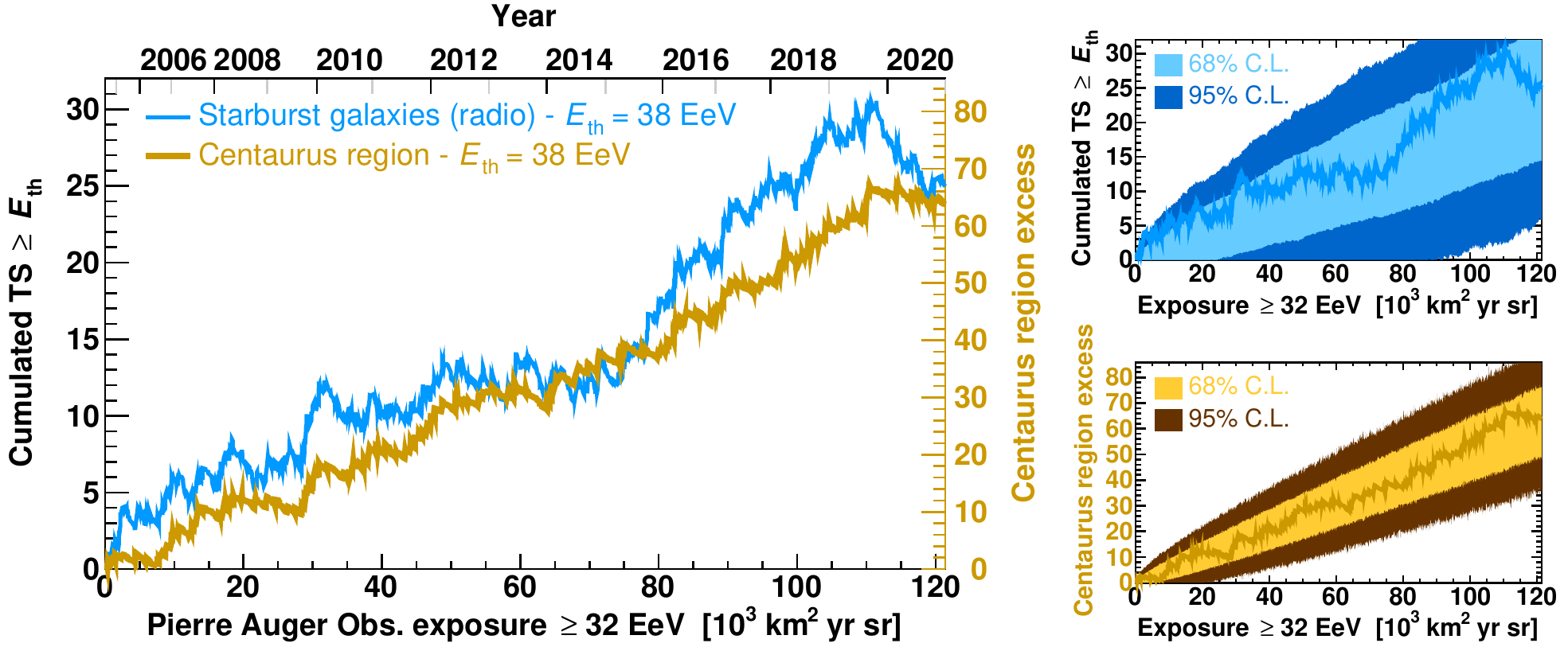}
\caption{Test statistic of the starburst model and excess in the Centaurus region above the best energy threshold as a function of exposure accumulated by the Pierre Auger Observatory. The fluctuations around the expected linear behavior are consistent with those expected from signal simulations, as illustrated in the right-most panels. \label{fig:growth}}
\end{figure*}

Although only pieces of evidence for anisotropy on intermediate angular scale can be claimed with the Phase~1 high-energy data set, the continued operation of the array may enable the reach of the $5\:\sigma$ discovery threshold. The latter corresponds to a post-trial $p$-value of $2.9\times10^{-7}$ or $5.7\times10^{-7}$ considering a search for both excesses and deficits (2-sided test) or just for excesses (1-sided test). The growth of the signal in the Centaurus region, quantified by the excess of events with respect to the isotropic expectation, and the growth of the test statistic of starburst model are displayed as a function of accumulated exposure in Figure~\ref{fig:growth}. These analyses yield post-trial significances of 3.9--4.2\:$\sigma$ for a 1- or 2-sided test applied to the Phase~1 high-energy data set. Both the test statistic and the excess of events are expected to grow linearly with exposure and the fluctuations observed around such a linear behavior are consistent with those expected from simulations. The model-independent search in the Centaurus region shows the smallest fluctuations and may be the most robust approach to forecasting the evolution of the signal. Assuming a fixed top-hat angular scale $\Psi = 27^\circ$ and a continued growth of the excess at a rate of $5.2\pm 1.2$ events per 10,000\:km$^2$\:yr\:sr, the $5\:\sigma$ (1-sided) discovery threshold would be expected for a total accumulated exposure of $165{,}000 \pm 15{,}000$\:km$^2$\:yr\:sr (68\% C.L.), which would be within reach by the end of 2025 (${\pm}\:2$ calendar years) adopting an approach similar to that developed in the present study.

\section{Conclusion}
\label{sec:discussion}

We have presented the measurement and analysis of arrival directions of the highest-energy events detected at the Pierre Auger Observatory during its first phase of operation. With a total of 2,635 UHECR events above 32\:EeV and accumulated exposure of 122,000 km$^2$\:sr\:yr, no indication for anisotropies on angular scales ranging from  one to thirty degrees emerges from auto-correlation studies or from blind searches over the entire sky. This lack of significant deviation from isotropy can be attributed \textit{a posteriori} to the small amplitude of the anisotropic signal evidenced here, to the vastness of the parameter space that has been probed, in addition to the limited number of events at the highest energies. More focused searches along the Galactic center and Galactic plane do not reveal any excesses. The flux along these structures and the associated statistical uncertainty are $\Phi_{\rm GC}({\geq}\:40\:\text{EeV}, \Psi=25^\circ) = (10.9\pm1.1)\times10^{-3} \:\text{km}^{-2}\:\text{yr}^{-1}\:\text{sr}^{-1}$ and $\Phi_{\rm GP}({\geq}\:40\:\text{EeV}, \Psi=25^\circ) = (9.8\pm 0.7)\times10^{-3} \:\text{km}^{-2}\:\text{yr}^{-1}\:\text{sr}^{-1}$, respectively. These values can be compared to average flux over the field of view of the Observatory $\Phi_{\rm ISO}({\geq}\:40\:\text{EeV}) = (11.3 \pm 0.4)\times10^{-3} \:\text{km}^{-2}\:\text{yr}^{-1}\:\text{sr}^{-1}$. A study along the supergalactic plane, not distinguishing among the various galaxies forming this structure, similarly yields $\Phi_{\rm SGP}({\geq}\:40\:\text{EeV}, \Psi=25^\circ)= (9.8\pm 0.6)\times10^{-3} \:\text{km}^{-2}\:\text{yr}^{-1}\:\text{sr}^{-1}$. 

Accounting for the attenuation of the UHECR mix inferred from lower energy observations, the sky viewed from the Pierre Auger Observatory is better modeled with a ${\sim}\:10\%$ flux excess in the directions of nearby galaxies observed in the radio, near-infrared, X-ray and gamma-ray bands. A 1-sided test for an excess disfavors isotropy at the $3.3-4.2\:\sigma$ level, depending on the catalog. A model-independent analysis of the Centaurus region, which contains the most prominent active and star-forming galaxies expected to contribute at these energies, reveals an excess that is significant at the $4.1\:\sigma$ C.L.

\begin{figure*}[t]
\epsscale{0.8}
\plotone{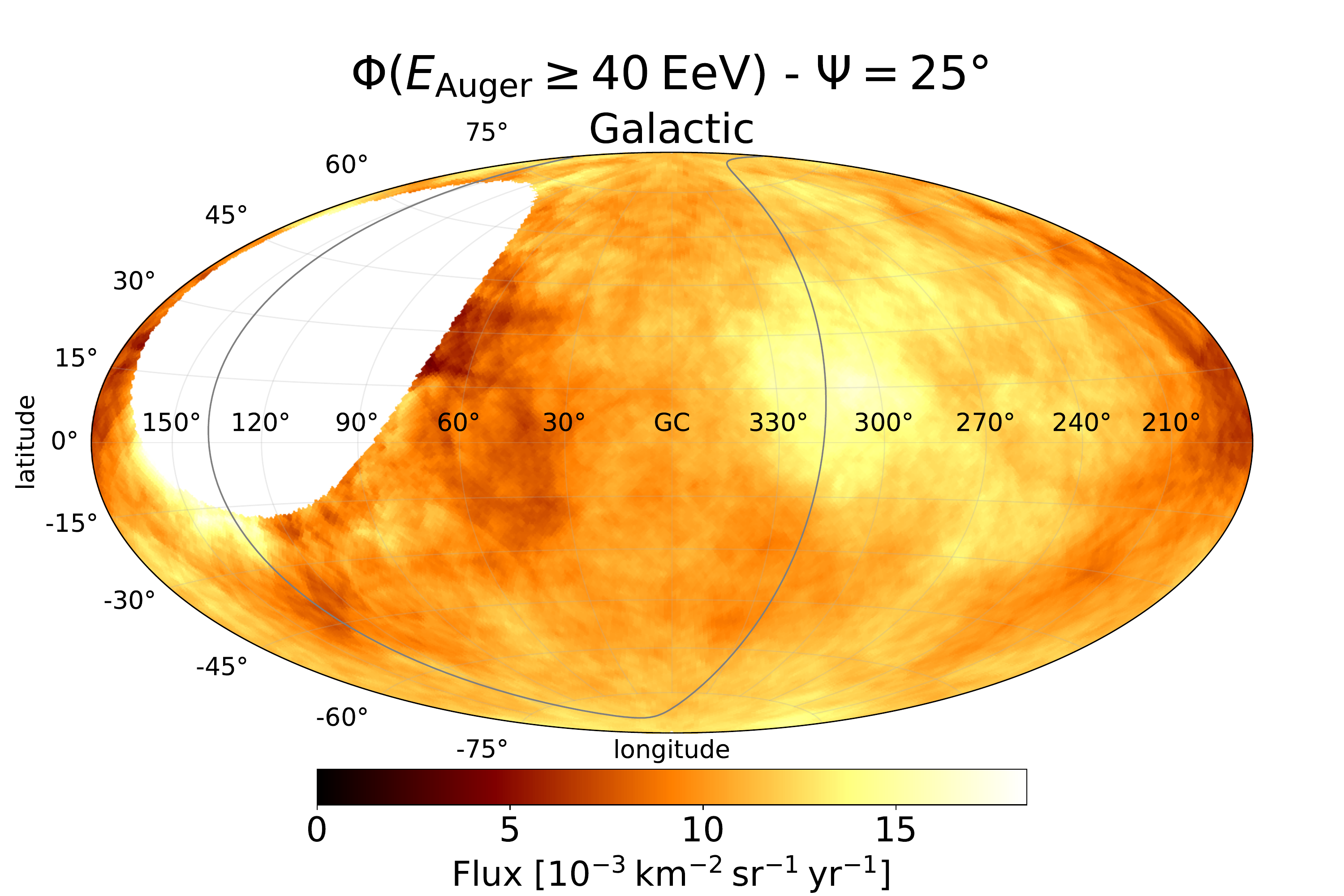}
\caption{Flux map at energies above 40\:EeV with a top-hat smoothing radius $\Psi = 25^\circ$ in Galactic coordinates. The supergalactic plane is shown as a gray line. The blank area is outside the field of view of the Pierre Auger Observatory. The complete Figure set (49 images), which shows the map as a function of energy threshold, is available in the online journal, in the arXiv source file and on the \augerwebsite\ of the Pierre Auger Collaboration. \label{fig:flux_map}
}
\end{figure*}

The average flux above 40\:EeV in a $25^\circ$ top-hat region centered on Centaurus~A can be estimated to $\Phi_{\rm Cen}({\geq}\:40\:\text{EeV}, \Psi=25^\circ) = (15.9\pm 1.3)\times10^{-3} \:\text{km}^{-2}\:\text{yr}^{-1}\:\text{sr}^{-1}$. In comparison, regions centered on the Virgo cluster and on the starburst galaxy NGC\:253 show fluxes of $\Phi_{\rm Virgo}({\geq}\:40\:\text{EeV}, \Psi=25^\circ) = (12.2\pm 1.8)\times10^{-3} \:\text{km}^{-2}\:\text{yr}^{-1}\:\text{sr}^{-1}$ and $\Phi_{\rm NGC\,253}({\geq}\:40\:\text{EeV}, \Psi=25^\circ) = (12.8\pm 1.2)\times10^{-3} \:\text{km}^{-2}\:\text{yr}^{-1}\:\text{sr}^{-1}$. As illustrated by the model sky maps in Appendix~\ref{app:catalogs}, the regions of NGC\:253 and of the Virgo cluster could be expected to be as bright as and brighter than the Centaurus region if the UHECR emission rate was simply traced by star-formation rate and stellar mass, respectively. At the present stage, although the starburst catalog enables the identification of the most significant deviation from isotropy (4.2\:$\sigma$) and the jetted AGN catalog the least significant deviation (3.3\:$\sigma$), no firm preference for correlation with a specific class of galaxies can be stated. It should further be noted that such a preferred correlation would not necessarily suggest causation in the form of the identification of the origin of UHECRs, as regular and turbulent magnetic fields traversed by these charged particles could alter the anisotropic pattern observed on Earth \cite[e.g.][]{2008PhRvD..77l3003K,2016APh....85...54E, 2019JCAP...05..004F,2022MNRAS.511..448B}.

Though the most significant deviation from isotropy is found at energies around ${\sim}\:40\:\text{EeV}$ for almost all the analyses, the excess is also hinted at for all catalogs and the Centaurus region at energies around ${\sim}\:60\:\text{EeV}$, as shown in Figure~\ref{fig:flux_map} (see online material). Indeed, it was in this higher energy range that the first indication of anisotropy was found in early Auger data \citep{2007Science.1151124}. An interpretation of the energy evolution of the signal on intermediate angular scales could be drawn in terms of maximum energy achieved for higher-charge nuclei. In a Peters' cycle scenario such as discussed in Section~\ref{sec:Cen}, the evidence for anisotropy above ${\sim}\:40\:$EeV would be interpreted as stemming from CNO nuclei, which would suggest $Z \approx 10-12$ nuclei to be responsible for the departure from isotropy above ${\sim}\:60\:$EeV. The estimate of maximum rigidity used here is based on the combined fit of spectrum and depth of shower maximum performed in \cite{2017JCAP...04..038A}. The direct inclusion in such analyses of arrival-direction information will enable us to test more directly this scenario. If this scenario of local extragalactic sources is extrapolated to lower energies, one could expect a contribution from He nuclei \citep[see e.g.][]{2009_Lemoine} in the energy range where a significant dipole, but no significant quadrupole has been reported using data from the Observatory. The strength of such an anisotropic contribution could nonetheless be further diluted in the contribution from more distant sources. 
We foresee that an in-depth comparison could be drawn studying the evolution of the large-scale dipolar and quadrupolar components as a function of energy.\footnote{We checked that no significant large-scale deviation from isotropy can be inferred from arrival-direction data in the energy range covered here, with constraints on the dipolar and quadrupolar components not in tension with those expected from best-fit catalog-based models \citep[as inferred e.g.\ for the 2MASS Redshift Survey in][]{2018_diMatteo_mnras}.} Alternatively, a more model-dependent but also more-constrained approach could exploit full-sky flux-limited catalogs encompassing galaxies out to the cosmic-ray horizon at the ankle energy.  

At this stage, it is not possible to make claims on which are the sources of the highest energy particles known in the Universe. This is in part due to the deflection they suffer in magnetic fields. Identifying the sources of UHECRs indeed runs parallel to deducing properties of Galactic and extragalactic magnetic fields, and constraints on one of these will enhance our understanding of the other. An important step will be taken through the inclusion of composition-sensitive observables in arrival direction studies. This will be done either through searches for anisotropy in the moments of such composition observables or by their use, event by event, to select only candidate light nuclei.
Future studies using the Observatory offer the promise to do so by means of the AugerPrime upgrade, currently being completed, which will enhance mass discrimination with the 100\% duty cycle of the surface detector.

\input{acknowledgments}


\appendix

\section{Data}
\label{app:data}

The data set used here consists of 2,635 events above 32\:EeV collected at the Pierre Auger Observatory from 1~January 2004 to 31~December 2020. The data set is formatted as shown in Table~\ref{tab:data}, which lists the twenty highest energy events. For each event, we report the year in which the event was detected, the Julian day of the year and the time of detection in UTC seconds. The arrival directions are expressed in local coordinates, $(\theta, \phi)$, the zenith and azimuth angle (measured counterclockwise from the east), respectively, and in equatorial coordinates (J2000), $(\alpha, \delta)$, the right ascension (R.A.) and declination (Dec), respectively. Finally, the reconstructed energy, in EeV, and the integrated exposure accumulated up to the time of detection are reported in the last two columns. The full list of 2,635 events, with the same information as in Table~\ref{tab:data}, is available at \ZenodoLink\ together with the code, the structure of which is described in Appendix~\ref{app:code}. 

The energies and arrival directions of the events may have changed with respect to those already released in previous works, such as \cite{2015ApJ...804...15A}. These changes are due to the refinements in the reconstruction reported in Section~\ref{sec:data} and to updates in the energy scale and calibration which were improved over the years. Similarly, a subsample of the vertical events used here is included in the recent Data Release from the Collaboration \citep{the_pierre_auger_collaboration_2021_4487613}. The latter were derived with the other reconstruction software used in the Collaboration, which enables independent cross checks and shows good consistency with the reconstruction software used here \citep{2020JInst..15P0021A}.

As mentioned in Section~\ref{sec:data}, the ratio of the number of inclined and vertical events is energy dependent. Anisotropy itself could impact the ratio of inclined and vertical events, as the two exposures differ over the sky. This effect is however small: the excess reported in Section~\ref{sec:Cen} would imply an expected ratio of $N_\text{\rm incl}/N_\text{\rm vert}=0.273$ instead of $0.278$ for an isotropic distribution. Above $32\:\text{EeV}$, a non-significant excess of inclined events is observed with respect to expectations from the exposure ratio and finite energy resolution ($N_{\rm incl}/N_{\rm vert}=0.292 \pm 0.014$). Above $80\:\text{EeV}$, there are 10 events with $\theta\geq60^\circ$ and 86 with $\theta<60^\circ$, which corresponds to a ratio of $N_\text{\rm incl}/N_\text{\rm vert}=0.116 \pm 0.039$. The deficit of inclined events is most significant above ${\sim}\:90\:$EeV, which results in a post-trial significance (under the assumption of isotropy) at the level of ${\sim}\: 2.5\: \sigma$, when penalized for a search as a function of energy. Such a discrepancy or a stronger one would have a 1.3\% probability of being found as a statistical fluctuation under the hypothesis that the energy calibrations of both data streams are correct. For completeness, we also consider the hypothesis that the deficit of inclined events at the highest energies is at least partly due to a systematic underestimation of inclined energies (or overestimation of vertical ones), as different reconstruction  techniques are used for the two sets. We tested for this effect empirically by selecting the events with zenith angles between $57^\circ < \theta < 63^\circ$ that are reconstructed by both the vertical and inclined reconstructions and for which six active stations surround the one closest to the core position. There are 161 such events and a power-law relation of the form $E_{\rm vert}=A\cdot E_{\rm incl}^B$ was fitted to extract the parameters $(A, B)$ that would convert the energies obtained from the inclined reconstruction to the energies obtained from the vertical reconstruction. The results are such that $E_{\rm vert} = 80\:\textrm{EeV}$ would correspond to $E_{\rm incl} = 76.1 \pm 1.6 \:\textrm{EeV}$. We applied the change to the energies of events in the inclined data set and performed, as a cross-check, the likelihood analysis with the starburst catalog (as in Section~\ref{sec:cat}) and the Centaurus-region analysis (as in Section~\ref{sec:Cen}). In both cases, we found the same results presented with the standard data set. This cross-check demonstrates that the possible systematic uncertainties induced by the difference in energy calibration of the vertical and inclined reconstructions do not affect the results presented in this paper.

\begin{deluxetable}{ccccccccc}[ht]
\tablecaption{Excerpt of the full data set of 2,635 events above 32\:EeV collected at the Pierre Auger Observatory between 1~January 2004 and 31~December 2020.\label{tab:data}}
\tablehead{\colhead{Year} &\colhead{JD}& \colhead{UTC} & \colhead{Zenith angle, $\theta$} & \colhead{Azimuth angle, $\phi$} & \colhead{R.A., $\alpha$} & \colhead{Dec, $\delta$} & \colhead{$E$}& \colhead{Cumulative exposure} \\ \colhead{$\mathrm{}$} & \colhead{$\mathrm{}$} & \colhead{$\mathrm{s}$} & \colhead{$\mathrm{{}^{\circ}}$} & \colhead{$\mathrm{{}^{\circ}}$} & \colhead{$\mathrm{{}^{\circ}}$} & \colhead{$\mathrm{{}^{\circ}}$} & \colhead{$\mathrm{EeV}$}  & \colhead{$\mathrm{km}^{2}\:\mathrm{sr}\:\mathrm{yr} $}} 
\startdata
2019 & 314 & 1573399408 & 58.6 &  $-$135.6 & 128.9 &  $-$52.0 & 166 & 111,900 \\
2007 & \phantom{0}13 & 1168768186 & 14.2 & \phantom{0$-$}85.6 & 192.9 &  $-$21.2 & 165 & \phantom{00}9,800 \\
2020 & 163 & 1591895321 & 18.9 &  \phantom{0}$-$47.7 & 107.2 &  $-$47.6 & 155 & 116,800 \\
2014 & 293 & 1413885674 & 6.8 &  $-$155.4 & 102.9 &  $-$37.8 & 155 & \phantom{0}70,600 \\
2018 & 224 & 1534096475 & 47.9 & \phantom{$-$}141.7 & 125.0 &  \phantom{0}$-$0.6 & 147 & 101,400 \\
2008 & 268 & 1222307719 & 49.8 & \phantom{$-$}140.5 & 287.8 & \phantom{0$-$}1.5 & 140 & \phantom{0}21,300 \\
2019 & 117 & 1556436334 & 14.8 &  \phantom{0}$-$32.7 & 275.0 &  $-$42.1 & 133 & 107,400 \\
2017 & 361 & 1514425553 & 41.7 &  \phantom{0}$-$30.5 & 107.8 &  $-$44.7 & 132 & \phantom{0}96,100 \\
2014 & \phantom{0}65 & 1394114269 & 58.5 & \phantom{0$-$}47.3 & 340.6 &  \phantom{$-$}12.0 & 131 & \phantom{0}65,300 \\
2005 & 186 & 1120579594 & 57.3 & \phantom{$-$}155.7 & \phantom{0}45.8 &   \phantom{0}$-$1.7 & 127 & \phantom{00}3,100 \\
2015 & 236 & 1440460829 & 20.1 &  \phantom{0}$-$46.1 & 284.8 &  $-$48.0 & 125 & \phantom{0}77,700 \\
2008 & \phantom{0}18 & 1200700649 & 50.3 & \phantom{$-$}178.9 & 352.5 &  $-$20.8 & 124 & \phantom{0}16,100 \\
2016 & \phantom{0}26 & 1453874568 & 22.6 &  \phantom{0}$-$14.7 & 175.6 &  $-$37.7 & 122 & \phantom{0}81,200 \\
2016 & \phantom{0}21 & 1453381745 & 13.7 &  $-$179.8 & 231.4 &  $-$34.0 & 122 & \phantom{0}81,100 \\
2011 & \phantom{0}26 & 1296108817 & 24.9 & \phantom{0$-$}90.9 & 150.0 &  $-$10.4 & 116 & \phantom{0}39,300 \\
2016 & \phantom{0}68 & 1457496302 & 23.7 & \phantom{$-$}108.7 & 151.5 &  $-$12.6 & 115 & \phantom{0}82,100 \\
2015 & 268 & 1443266386 & 77.2 &  $-$172.0 & \phantom{0}21.7 &  $-$13.8 & 113 & \phantom{0}78,400 \\
2016 & 297 & 1477276760 & 49.5 & \phantom{$-$}104.5 & 352.1 &  \phantom{$-$}13.2 & 111 & \phantom{0}86,800 \\
2020 & \phantom{0}66 & 1583535647 & 41.4 & \phantom{0}$-$20.6 & 133.6 &  $-$38.3 & 110 & 114,600 \\
2018 & 174 & 1529810463 & 42.7 & \phantom{$-$00}4.3 & 300.0 & $-$22.6 & 110 & 100,200 \\
\enddata
\tablecomments{See text for a description of the columns. Events are sorted here by decreasing energy, $E$, and only the 20 highest-energy events are displayed. The full data set is available in the same format at \ZenodoLink.}
\end{deluxetable}

\clearpage
\section{Code}
\label{app:code}

\begin{figure*}[ht]
\includegraphics[width=\textwidth]{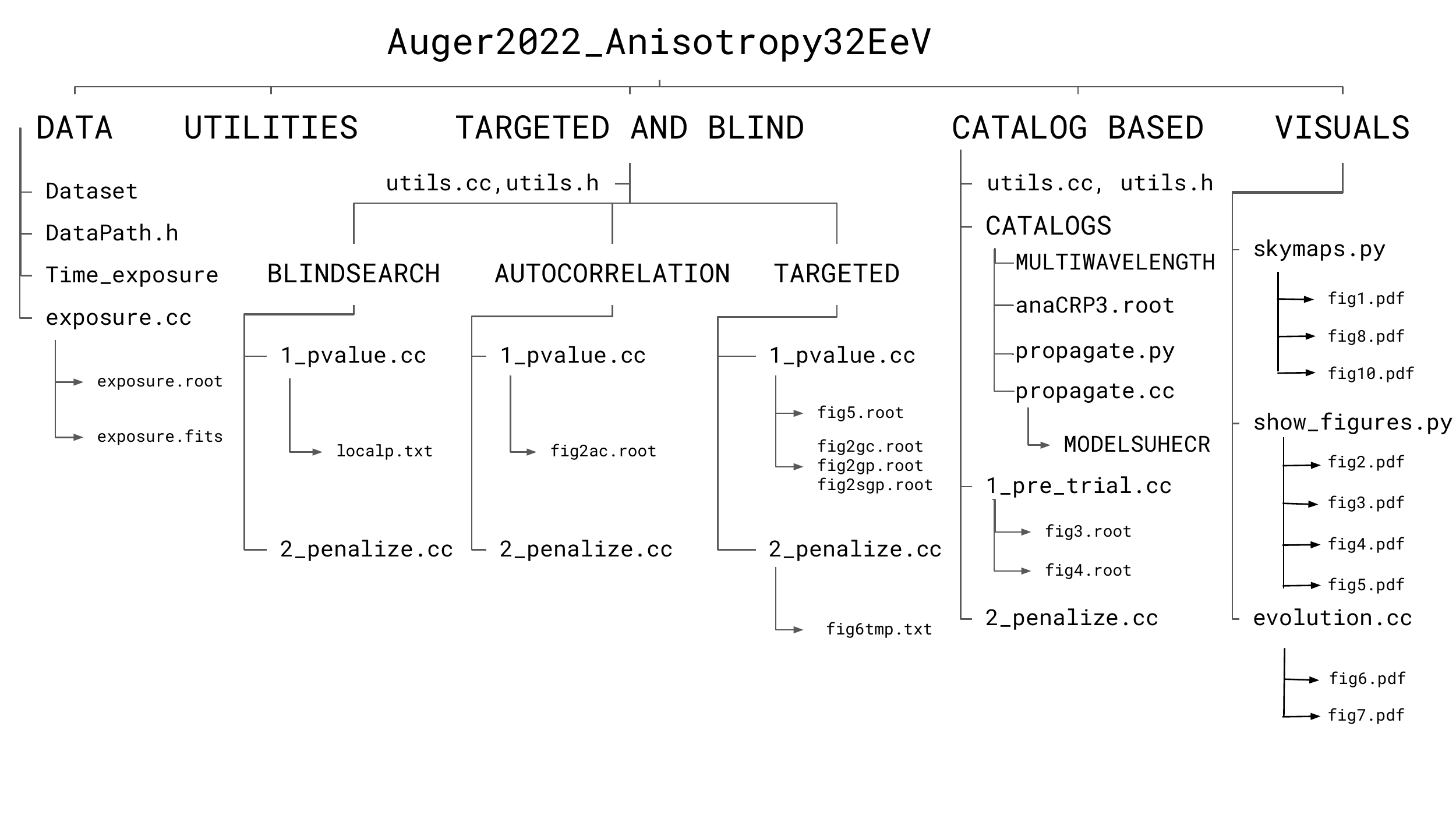}
\vspace{-1cm}
\caption{Schematic view of the code. \label{fig:code}}
\end{figure*}

The structure of the code used to produce the results of this paper is presented in Figure~\ref{fig:code}. The main analyses are contained in two folders, called \texttt{Targeted\_Blind} for Sections~\ref{sec:blind} and \ref{sec:Cen}, and \texttt{Catalog\_Based} for Section~\ref{sec:likelihood}. The add-ons and utilities needed to run the analyses are contained in the folders \texttt{Data}, \texttt{Utilities} and \texttt{Visuals}. A brief description of each folder follows.

\begin{itemize}

\item The \texttt{Data} folder contains the file of the data set used in all of the analyses, named \texttt{AugerApJS2022\_Yr\_JD\_UTC\_Th\_Ph\_RA\_Dec\_E\_Expo.dat}. Additionally, the folder contains a C++ script named \texttt{exposure.cc}, which computes the directional exposure of the Observatory for both vertical and inclined events, which is integrated over the duration of the acquisition period; the exposure script produces two files, \texttt{exposure.root} and \texttt{exposure.fits}, which contain the exposure as a function of declination in TF1-root format and in healpixmap-fits form (RING scheme, Galactic coordinates), respectively. The \texttt{Time\_exposure} file provides the evolution of exposure with time, as displayed in the upper axis of Figure~\ref{fig:growth}. The \texttt{DataPath.h} contains the declaration of the data set file to be used by all the analyses for easy user intervention.

\item The \texttt{Utilities} folder contains files with auxiliary classes and functions used by all other parts of the code, in particular coordinate-conversion utilities and \texttt{HEALPix} map manipulation.

\item The \texttt{Targeted\_Blind} folder contains the code for the targeted (Sections~\ref{sec:Cen} and \ref{sec:struc}), blind (Section~\ref{sec:locXS}) and autocorrelation (Section~\ref{sec:autoc}) analyses. The first folder level contains:
\begin{itemize}
    \item dedicated utilities (\texttt{utils.h} and \texttt{utils.cc});
    \item three sub-folders: \texttt{Blind}, \texttt{Targeted} and \texttt{Autocorrelation} containing the respective analyses. 
\end{itemize}
Each of the three sub-folders contains a script performing the computation of the local $p$-value, \texttt{1\_pvalue.cc}, and a script penalizing for the search over the parameter space, \texttt{2\_penalization.cc}. The results of the \texttt{Blind} code are stored in \texttt{.txt} files for easy readout, while the results of \texttt{Autocorrelation} code are stored as a \texttt{.root} file. In the \texttt{Targeted} folder, \texttt{1\_pvalue.cc} produces the outputs \texttt{fig5.root}, \texttt{fig2gc.root}, \texttt{fig2gp.root} and \texttt{fig2sgp.root}, which contain the local $p$-value in bins of energy threshold and search radius for the Centaurus region, Galactic center, Galactic plane and supergalactic plane analyses, respectively; the \texttt{.root} files in the \texttt{Autocorrelation} and \texttt{Targeted} folders contain also copies of the support histograms used to calculate the local $p$-value which are used by the penalization algorithm. The script \texttt{2\_penalization.cc} produces the post-trial $p$-value in single-value text form, as well as the file \texttt{fig6tmp.txt}, which stores the pre-trial $p$-values as a function of energy threshold penalized only for the scan in angle (see Figure~\ref{fig:Edependence}).

\item The \texttt{Catalog\_Based} folder contains the code for the likelihood-ratio analysis. The first folder level comprises: 
\begin{itemize}
    \item dedicated utilities (\texttt{utils.h} and \texttt{utils.cc});
    \item the folder \texttt{Catalogs}, which contains the raw catalogs of galaxies in the subfolder \texttt{Multiwavelength}, as described in Appendix~\ref{app:catalogs}. The raw catalog files are input, above a fixed energy threshold, to the script \texttt{propagate.cc}, in conjunction with the Auger composition model contained in the file \texttt{AnaCRP3.root}, to produce the attenuated models used in the analysis. These attenuated models can be produced above all energy thresholds by running the script  \texttt{propagate.py}, with outputs stored in the subfolder \texttt{ModelsUHECR}. The latter is organized as different folders for each catalog;
    \item the analysis routines: \texttt{1\_pre\_trial.cc}, which produces the results stored in the files \texttt{fig3.root}, showing the test statistic, signal fraction and search radius as a function of threshold energy, and \texttt{fig4.root}, showing the test statistic as a function of the signal fraction and search radius with 68$\%$ C.L. contours for each catalog; \texttt{2\_penalization.cc} produces the post-trial $p$-values.
\end{itemize} 

\item The \texttt{Visuals} folder contains scripts that produce the figures shown in the paper. The python script \texttt{skymaps.py} produces the sky maps in Hammer-Aitoff view: the Li-Ma significance map, \texttt{fig1.pdf}, the flux maps above successive energy threshold stored in \texttt{fig8.pdf} and the model maps stored in \texttt{fig10.pdf}. The script \texttt{show\_figures.py} produces \texttt{fig2.pdf}, \texttt{fig3.pdf}, \texttt{fig4.pdf} and \texttt{fig5.pdf} from their respective root files. The script \texttt{evolution.cc} produces \texttt{fig6.pdf}, the plot of the pre-trial $p$-values from the Centaurus-region analysis and likelihood-ratio analysis against starburst galaxies as a function of threshold energy; it also produces \texttt{fig7.pdf}, the plot of the evolution of test statistic of the starburst analysis and of the excess in the Centaurus region as a function of the exposure accumulated at the Observatory.
\end{itemize}

\clearpage
\section{Catalogs}
\label{app:catalogs}

\begin{figure}[ht]
\epsscale{1.15}
\plottwo{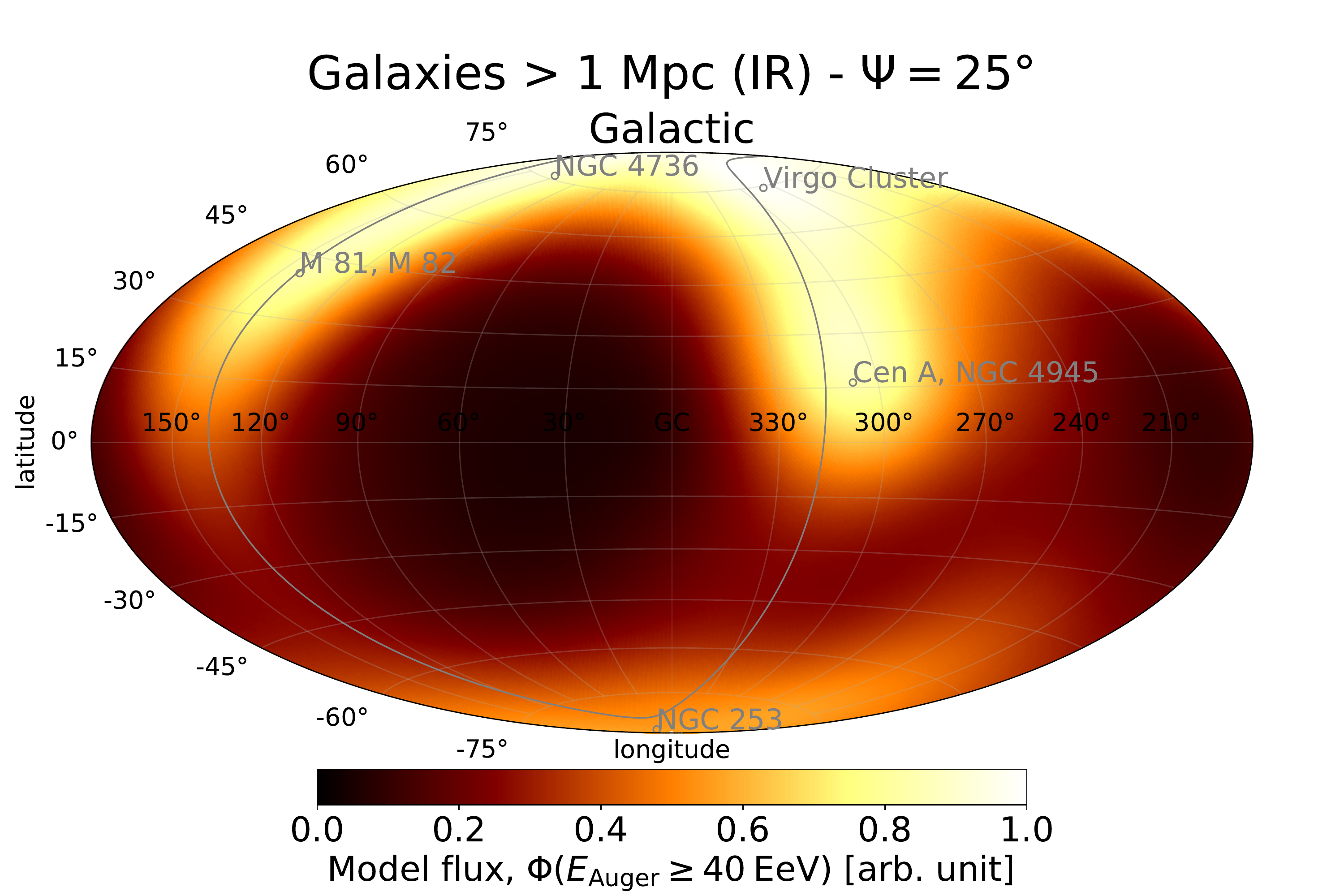}{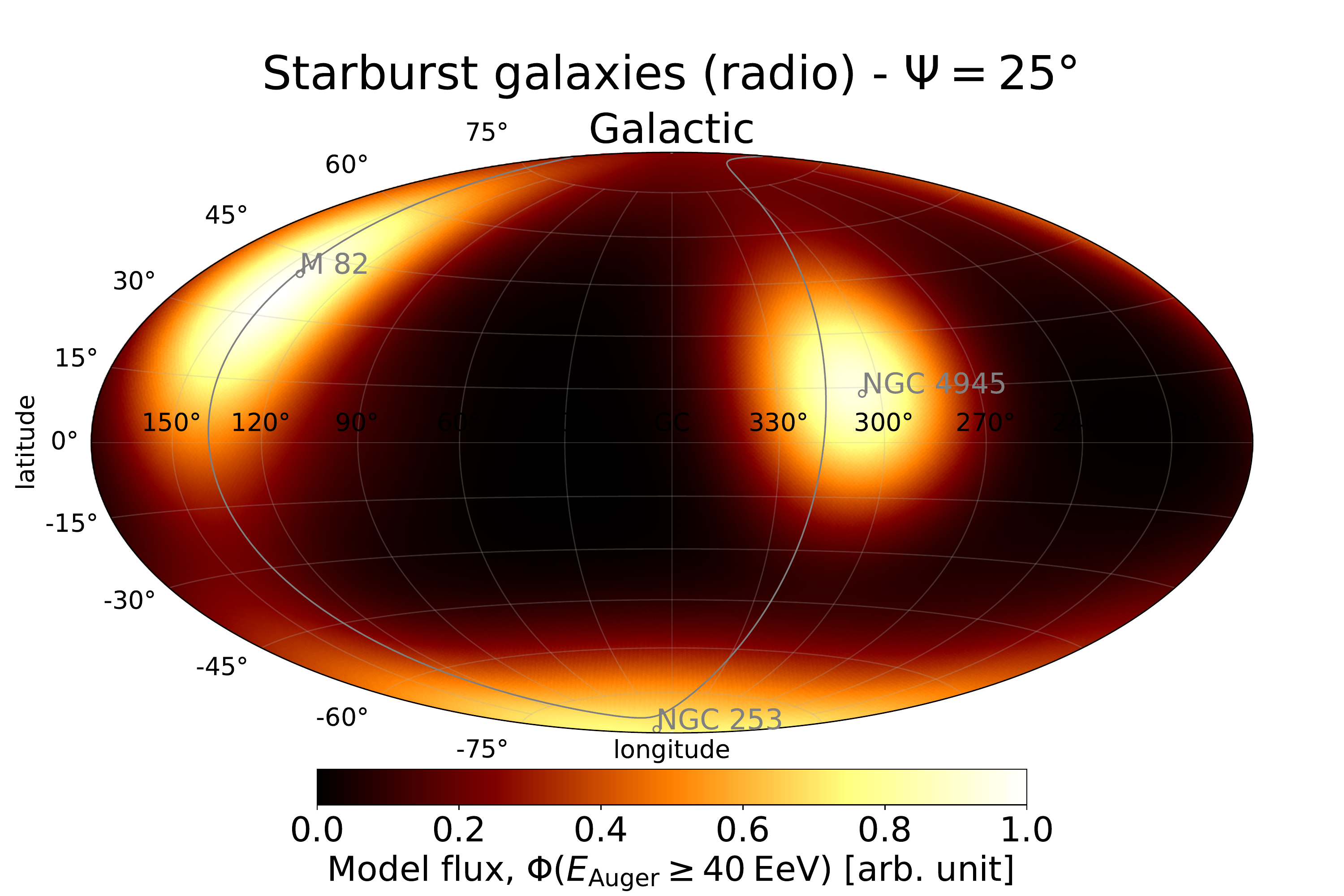}
\plottwo{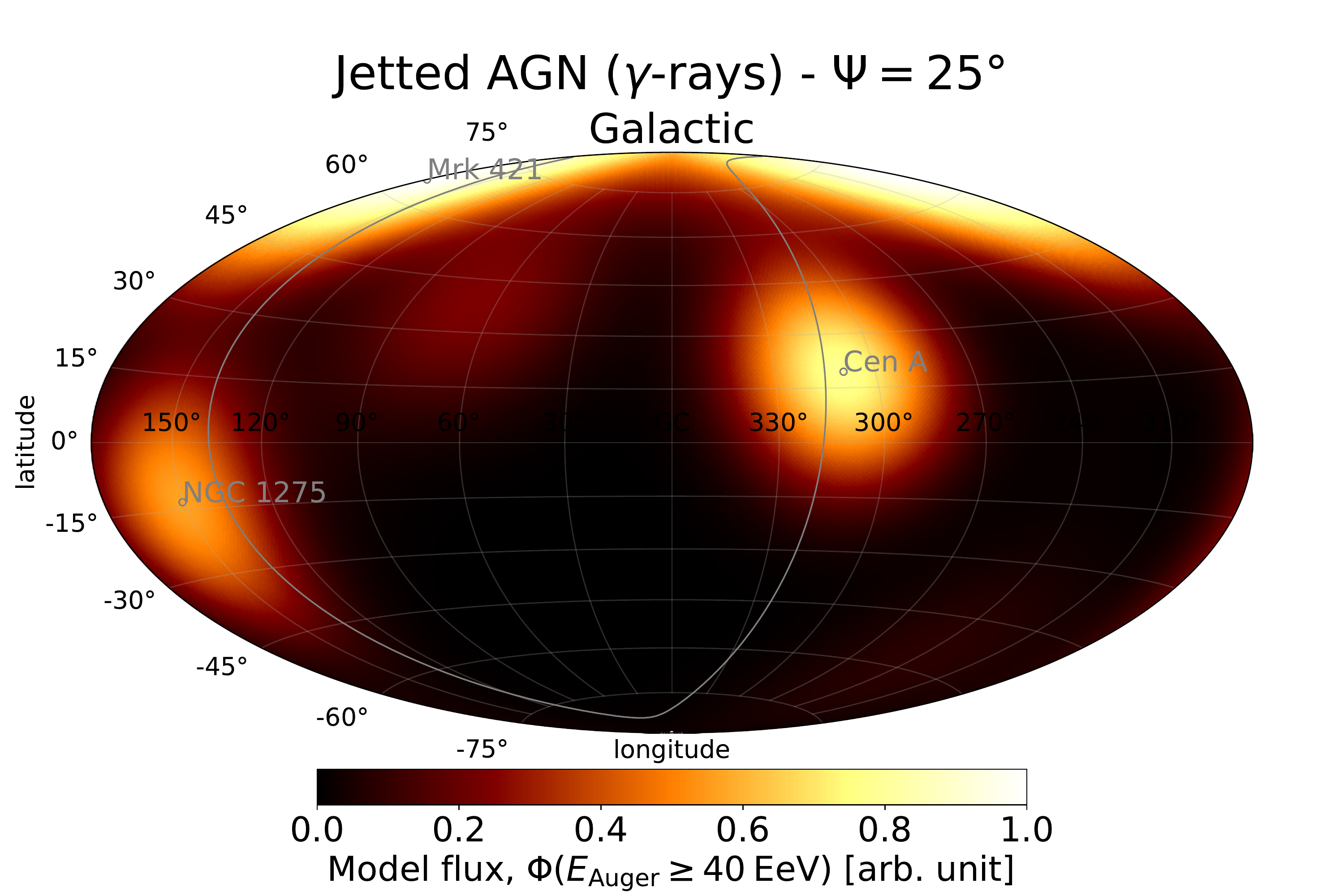}{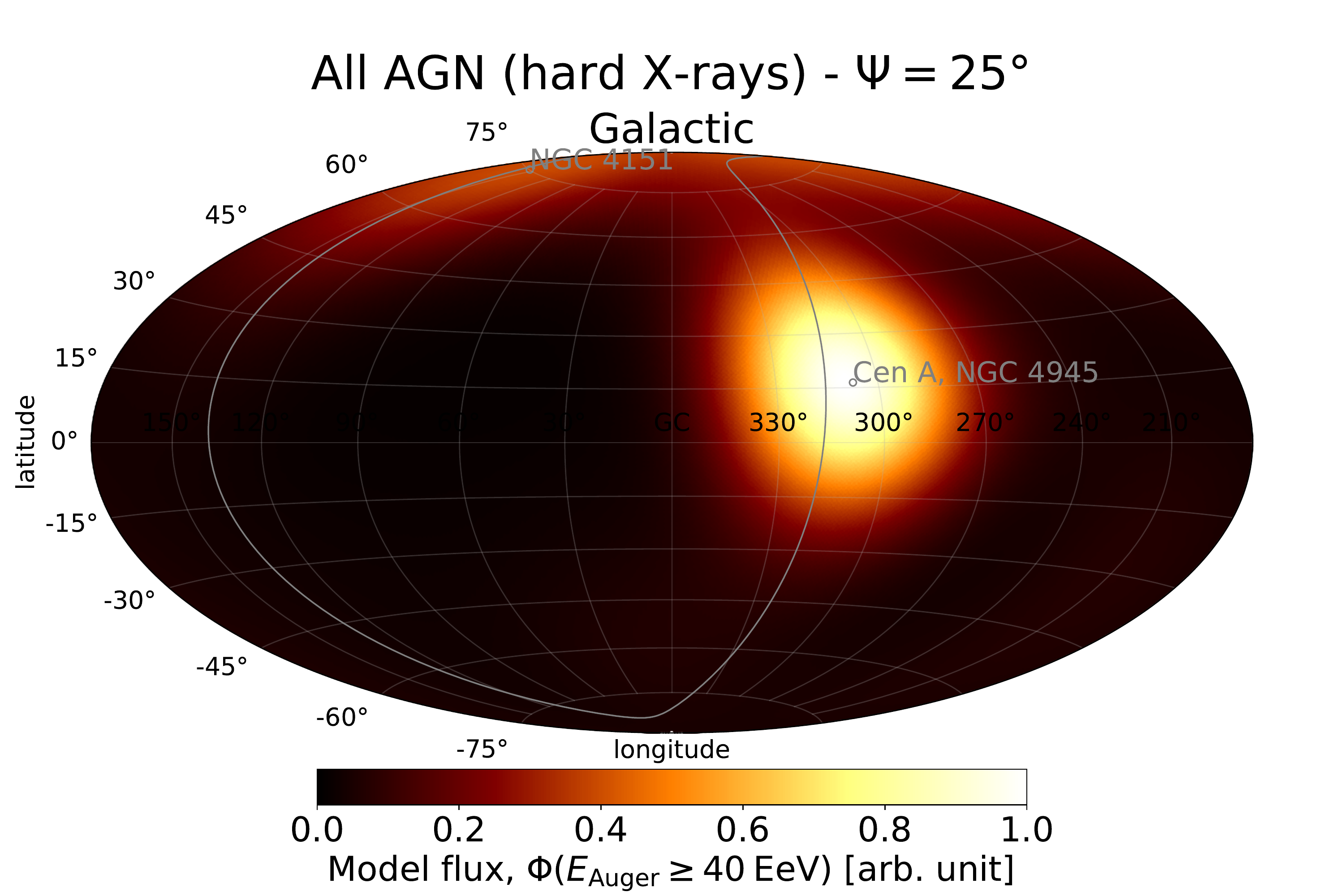}
\caption{Best-fit UHECR source models above 40\:EeV with a top-hat smoothing radius $\Psi = 25^\circ$ in Galactic coordinates. The supergalactic plane is shown as a gray line. Prominent sources in each of the catalogs are marked with gray circles. \label{fig:model_maps}}\vspace{-0.25cm}
\end{figure}

The best-fit sky models above 40\:EeV obtained with the four catalogs described in Section~\ref{sec:cat} are shown in Figure~\ref{fig:model_maps}. These sky maps do not include any isotropic component and display only the flux expected from galaxies included in the catalogs, which is smeared on the best-fit Fisher angular scale above 40\:EeV obtained with each catalog. A further top-hat smoothing on an angular scale $\Psi=25^\circ$ is performed for the sake of comparison with Figure~\ref{fig:flux_map}.\vspace{-0.25cm}

The models shown in Figure~\ref{fig:model_maps} are based on the UHECR flux expected from each galaxy in proportion to its electromagnetic flux. The multiwavelength information on the galaxies is made available in the \texttt{Multiwavelength} sub-folder of the catalog-based study, as described in Appendix~\ref{app:code}, and is available online at \ZenodoLink. The \texttt{Multiwavelength} folder contains one file per catalog, with tabulated values detailed in Tables~\ref{tab:2mass}, \ref{tab:sbg}, \ref{tab:bat} and \ref{tab:gagn}. The first column of each of these tables provides the name of the source as referenced by the authors of the source catalog. The second column provides a counterpart name that is consistent across all four catalogs. The third column provides the type of galaxy, extracted either from the source catalog or from the HyperLEDA database. The fourth and fifth columns provide the equatorial coordinates of the galaxy. The sixth and seventh columns display the distance modulus and associated uncertainty extracted from the \texttt{modbest} entry of the HyperLEDA database. The eighth and ninth columns display the corresponding luminosity distance in Mpc as well as the relative uncertainty on this quantity. The electromagnetic flux of each galaxy is provided in column 10, except in Table~\ref{tab:2mass} where the K-band magnitude is provided. Whenever available, the uncertainty on the quantity provided in column 10 is shown in column 11. Finally, a flag is provided in the last columns of Tables~\ref{tab:sbg}, \ref{tab:bat} and \ref{tab:gagn}. This flag indicates whether the galaxy was also included in the main samples studied in \cite{2018ApJ...853L..29A} (Y), in one of the cross-check samples (X), or not included in earlier versions of these catalogs (N). The flag column of Table~\ref{tab:bat} indicates the origin of the redshift estimate, either from HyperLEDA or from NED for the 23 X-ray AGNs that are not listed in HyperLEDA.

\begin{changemargin}{24cm}{1cm}

\begin{longrotatetable}
\begin{footnotesize}

\begin{deluxetable}{ccccccccccc}
\tablecaption{Galaxies (2MASS(K$<$11.75) $\times$ HyperLEDA).\label{tab:2mass}}
\tablehead{\colhead{PGC} & \colhead{Counterpart} & \colhead{Object Type} & \colhead{R.A.} & \colhead{Dec} & \colhead{$(m-M)$} & \colhead{$\sigma(m-M)$} & \colhead{$d_{\rm L}$} & \colhead{$\sigma(d_{\rm L})/d_{\rm L}$} & \colhead{$\rm K_t$} & \colhead{$\sigma({\rm K_t})$}\\ \colhead{$\mathrm{}$} & \colhead{$\mathrm{}$} & \colhead{$\mathrm{}$} & \colhead{$\mathrm{{}^{\circ}}$} & \colhead{$\mathrm{{}^{\circ}}$} & \colhead{$\mathrm{mag}$} & \colhead{$\mathrm{mag}$} & \colhead{$\mathrm{Mpc}$} & \colhead{$\mathrm{}$} & \colhead{$\mathrm{mag}$} & \colhead{$\mathrm{mag}$}}
\startdata
29128 & NGC3109     & G & 150.78 & -26.16 & 25.56 & 0.02 & 1.29 & 0.007 & 9.57 & 0.40 \\
29653 & PGC029653   & G & 152.75 & -4.69 & 25.59 & 0.03 & 1.31 & 0.013 & 11.31 & 0.56 \\
28913 & UGC05373    & G & 150.00 & 5.33 & 25.79 & 0.01 & 1.44 & 0.006 & 10.76 & 0.23 \\
100169 & PGC100169  & G & 31.52 & 69.00 & 26.15 & 0.20 & 1.70 & 0.092 & 9.69 & 0.24 \\
67908 & IC5152      & G & 330.67 & -51.30 & 26.46 & 0.03 & 1.96 & 0.012 & 9.05 & 0.36 \\
3238 & NGC0300      & G & 13.72 & -37.68 & 26.53 & 0.02 & 2.03 & 0.007 & 6.58 & 0.36 \\
1014 & NGC0055      & G & 3.72 & -39.20 & 26.62 & 0.01 & 2.11 & 0.006 & 6.34 & 0.18 \\
9140 & PGC009140    & G & 36.18 & -73.51 & 26.63 & 0.07 & 2.12 & 0.032 & 10.83 & 0.10 \\
13115 & UGC02773    & G & 53.03 & 47.79 & 26.69 & 0.20 & 2.18 & 0.092 & 9.80 & 0.10 \\
39573 & IC3104      & G & 184.69 & -79.73 & 26.86 & 0.02 & 2.36 & 0.007 & 9.24 & 0.14 \\
60849 & IC4662      & G & 266.79 & -64.64 & 27.03 & 0.01 & 2.55 & 0.006 & 9.45 & 0.21 \\
47495 & UGC08508    & G & 202.68 & 54.91 & 27.07 & 0.02 & 2.60 & 0.011 & 11.51 & 0.10 \\
40904 & UGC07577    & G & 186.92 & 43.50 & 27.08 & 0.02 & 2.60 & 0.011 & 10.45 & 0.20 \\
54392 & ESO274-001  & G & 228.56 & -46.81 & 27.24 & 0.06 & 2.80 & 0.026 & 8.30 & 0.39 \\
51472 & UGC09240    & G & 216.18 & 44.53 & 27.25 & 0.02 & 2.82 & 0.008 & 10.89 & 0.13 \\
39023 & NGC4190     & G & 183.44 & 36.63 & 27.26 & 0.04 & 2.83 & 0.020 & 11.40 & 0.77 \\
14241 & PGC014241   & G & 59.96 & 67.14 & 27.37 & 0.03 & 2.98 & 0.012 & 8.24 & 0.16 \\
4126 & NGC0404      & G & 17.36 & 35.72 & 27.37 & 0.02 & 2.98 & 0.007 & 7.53 & 0.02 \\
39225 & NGC4214     & G & 183.91 & 36.33 & 27.37 & 0.01 & 2.98 & 0.002 & 8.09 & 0.21 \\
38881 & NGC4163     & G & 183.04 & 36.17 & 27.38 & 0.02 & 2.99 & 0.007 & 10.92 & 0.08 \\
15488 & NGC1560     & G & 68.20 & 71.88 & 27.38 & 0.10 & 2.99 & 0.046 & 9.07 & 0.22 \\
49050 & ESO383-087  & G & 207.32 & -36.06 & 27.52 & 0.02 & 3.19 & 0.007 & 9.91 & 0.14 \\
15439 & PGC015439   & G & 68.01 & 63.62 & 27.53 & 0.05 & 3.20 & 0.024 & 10.97 & 0.17 \\
21396 & NGC2403     & G & 114.21 & 65.60 & 27.53 & 0.01 & 3.20 & 0.004 & 6.24 & 0.14 \\
47762 & NGC5206     & G & 203.43 & -48.15 & 27.53 & 0.01 & 3.21 & 0.005 & 8.39 & 0.25 \\
\dots & \dots & \dots & \dots & \dots & \dots & \dots & \dots & \dots & \dots & \dots \\
127001 & PGC127001  & G & 67.39 & -61.25 & 36.99 & 0.07 & 249.7 & 0.030 & 11.72 & 0.18
\enddata
\tablecomments{44,113 entries within 250\:Mpc. 17,143 entries at $d_{\rm L}< 100\:$Mpc, 39,563 at $d_{\rm L}< 200\:$Mpc.}
\end{deluxetable}

\end{footnotesize}
\end{longrotatetable}

\begin{longrotatetable}
\begin{footnotesize}

\begin{deluxetable}{cccccccccccc}
\tablecaption{Starburst galaxies (Lunardini+ '19).\label{tab:sbg}}
\tablehead{\colhead{Lunardi Name} & \colhead{Counterpart} & \colhead{Host Type} & \colhead{R.A.} & \colhead{Dec} & \colhead{$(m-M)$} & \colhead{$\sigma(m-M)$} & \colhead{$d_{\rm L}$} & \colhead{$\sigma(d_{\rm L})/d_{\rm L}$} & \colhead{$\Phi$(1.4\:GHz)} & \colhead{$\sigma(\Phi)$} & \colhead{flag: in Pierre Auger Collaboration 2018b?}\\ \colhead{$\mathrm{}$} & \colhead{$\mathrm{}$} & \colhead{$\mathrm{}$} & \colhead{$\mathrm{{}^{\circ}}$} & \colhead{$\mathrm{{}^{\circ}}$} & \colhead{$\mathrm{mag}$} & \colhead{$\mathrm{mag}$} & \colhead{$\mathrm{Mpc}$} & \colhead{$\mathrm{}$} & \colhead{$\mathrm{Jy}$} & \colhead{$\mathrm{Jy}$} & \colhead{(No/Yes/Xcheck)}}
\startdata
NGC0055 & NGC0055   & SBm & 3.72 & -39.20 & 26.62 & 0.01 & 2.11 & 0.005 & 0.37 & N/A & N \\
NGC1569 & NGC1569   & IB & 67.70 & 64.85 & 27.53 & 0.05 & 3.21 & 0.023 & 0.40 & N/A & X \\
NGC2403 & NGC2403   & SABc & 114.21 & 65.60 & 27.53 & 0.01 & 3.21 & 0.005 & 0.39 & N/A & X \\
IC342 & IC342       & SABc & 56.70 & 68.10 & 27.68 & 0.03 & 3.44 & 0.014 & 2.25 & N/A & Y \\
NGC4945 & NGC4945   & Sbc & 196.37 & -49.47 & 27.7 & 0.02 & 3.47 & 0.009 & 6.60 & N/A & Y \\
NGC3034(M82) & M82  & S? & 148.97 & 69.68 & 27.79 & 0.01 & 3.61 & 0.005 & 7.29 & N/A & Y \\
NGC0253 & NGC253    & SABc & 11.89 & -25.29 & 27.84 & 0.02 & 3.70 & 0.009 & 6.00 & N/A & Y \\
N/A & Circinus      & Sb & 213.29 & -65.34 & 28.12 & 0.36 & 4.21 & 0.166 & 1.50 & N/A & Y \\
NGC5236(M83) & M83  & Sc & 204.25 & -29.87 & 28.45 & 0.02 & 4.90 & 0.009 & 2.44 & N/A & Y \\
Maffei2 & Maffei2   & Sbc & 40.48 & 59.60 & 28.79 & 0.12 & 5.73 & 0.055 & 1.01 & N/A & X \\
NGC6946 & NGC6946   & SABc & 308.72 & 60.15 & 29.14 & 0.05 & 6.73 & 0.023 & 1.40 & N/A & Y \\
NGC4631 & NGC4631   & SBcd & 190.53 & 32.54 & 29.33 & 0.02 & 7.35 & 0.009 & 1.12 & N/A & Y \\
NGC5194(M51) & M51  & SABb & 202.48 & 47.20 & 29.67 & 0.02 & 8.59 & 0.009 & 1.31 & N/A & Y \\
NGC5055(M63) & NGC5055 & Sbc & 198.96 & 42.03 & 29.78 & 0.01 & 9.04 & 0.005 & 0.35 & N/A & Y \\
NGC2903 & NGC2903   & Sbc & 143.04 & 21.50 & 29.85 & 0.11 & 9.33 & 0.051 & 0.44 & N/A & Y \\
NGC891 & NGC891     & Sb & 35.64 & 42.35 & 29.94 & 1.72 & 9.73 & 0.792 & 0.70 & N/A & Y \\
NGC1068 & NGC1068   & Sb & 40.66 & 0.00 & 30.12 & 0.34 & 10.6 & 0.157 & 4.85 & N/A & Y \\
NGC3628 & NGC3628   & SBb & 170.07 & 13.59 & 30.21 & 0.34 & 11.0 & 0.157 & 0.47 & N/A & Y \\
NGC4818 & NGC4818   & SABa & 194.20 & -8.53 & 30.27 & 0.33 & 11.3 & 0.152 & 0.45 & N/A & N \\
NGC3627 & NGC3627   & Sb & 170.06 & 12.99 & 30.30 & 0.04 & 11.5 & 0.018 & 0.46 & N/A & Y \\
NGC1808 & NGC1808   & Sa & 76.93 & -37.51 & 30.45 & 0.36 & 12.3 & 0.166 & 0.50 & N/A & X \\
NGC4303 & M61       & Sbc & 185.48 & 4.47 & 30.45 & 0.10 & 12.3 & 0.046 & 0.44 & N/A & X \\
NGC3521 & NGC3521   & SABb & 166.45 & -0.04 & 30.47 & 0.29 & 12.4 & 0.134 & 0.35 & N/A & N \\
NGC0660 & NGC660    & Sa & 25.76 & 13.65 & 30.50 & 1.31 & 12.6 & 0.603 & 0.37 & N/A & Y \\
NGC4254 & NGC4254   & Sc & 184.71 & 14.42 & 30.77 & 1.13 & 14.3 & 0.520 & 0.37 & N/A & N \\
\dots & \dots & \dots & \dots & \dots & \dots & \dots & \dots & \dots & \dots & \dots & \dots \\
NGC6240 & NGC6240   & S0-a & 253.26 & 2.40 & 35.18 & 0.15 & 108.6 & 0.069 & 0.65 & N/A & Y
\enddata
\tablecomments{44 entries within 250\:Mpc. 43 entries at $d_{\rm L}< 100\:$Mpc, 44 at $d_{\rm L}< 200\:$Mpc.}
\end{deluxetable}

\end{footnotesize}
\end{longrotatetable}

\begin{longrotatetable}
\begin{footnotesize}

\begin{deluxetable}{cccccccccccc}
\tablecaption{Jetted and non-jetted AGNs (\textit{Swift}-BAT 105 months).\label{tab:bat}}
\tablehead{\colhead{BAT105 Name} & \colhead{Counterpart} & \colhead{AGN Type} & \colhead{R.A.} & \colhead{Dec} & \colhead{$(m-M)$} & \colhead{$\sigma(m-M)$} & \colhead{$d_{\rm L}$} & \colhead{$\sigma(d_{\rm L})/d_{\rm L}$} & \colhead{$\Phi$($14-195\:$keV)} & \colhead{$\sigma(\Phi)$} & \colhead{flag: ref. $(m-M)$}\\ \colhead{$\mathrm{}$} & \colhead{$\mathrm{}$} & \colhead{$\mathrm{}$} & \colhead{$\mathrm{{}^{\circ}}$} & \colhead{$\mathrm{{}^{\circ}}$} & \colhead{$\mathrm{mag}$} & \colhead{$\mathrm{mag}$} & \colhead{$\mathrm{Mpc}$} & \colhead{$\mathrm{}$} & \colhead{$10^{-12}\:$erg\:cm$^{-2}\:$s$^{-1}$} & \colhead{$10^{-12}\:$erg\:cm$^{-2}\:$s$^{-1}$} & \colhead{(HyperLEDA/NED)}}
\startdata
J1305.4-4928 & NGC4945  & Sy2       & 196.37 & -49.47 & 27.70 & 0.02 & 3.47 & 0.009 & 282.1 & N/A & H \\
J0955.5+6907 & M81      & Sy1.9     & 148.94 & 69.06 & 27.78 & 0.01 & 3.60 & 0.005 & 20.3 & N/A & H \\
J1325.4-4301 & CenA     & BeamedAGN & 201.37 & -43.02 & 27.83 & 0.03 & 3.68 & 0.014 & 1346.3 & N/A & H \\
J1412.9-6522 & Circinus & Sy2       & 213.29 & -65.34 & 28.12 & 0.36 & 4.21 & 0.166 & 273.2 & N/A & H \\
J1210.5+3924 & NGC4151  & Sy1.5     & 182.64 & 39.41 & 28.39 & 1.65 & 4.76 & 0.760 & 618.9 & N/A & H \\
J1202.5+3332 & NGC4395  & Sy2       & 186.45 & 33.53 & 28.39 & 0.01 & 4.76 & 0.005 & 27.5 & N/A & H \\
J0420.0-5457 & NGC1566  & Sy1.5     & 64.96 & -54.94 & 29.13 & 1.16 & 6.70 & 0.534 & 19.5 & N/A & H \\
J1219.4+4720 & M106     & Sy1.9     & 184.75 & 47.29 & 29.41 & 0.01 & 7.62 & 0.005 & 23.0 & N/A & H \\
J1329.9+4719 & M51      & Sy2       & 202.48 & 47.20 & 29.67 & 0.02 & 8.59 & 0.009 & 13.3 & N/A & H \\
J0242.6+0000 & NGC1068  & Sy1.9     & 40.66 & 0.00 & 30.12 & 0.34 & 10.6 & 0.157 & 37.9 & N/A & H \\
J1717.1-6249 & NGC6300  & Sy2       & 259.25 & -62.83 & 30.15 & 0.09 & 10.7 & 0.041 & 96.4 & N/A & H \\
J1203.0+4433 & NGC4051  & Sy1.5     & 180.78 & 44.52 & 30.28 & 0.35 & 11.4 & 0.161 & 42.5 & N/A & H \\
J1652.0-5915B & NGC6221 & Sy2       & 253.18 & -59.23 & 30.34 & 0.62 & 11.7 & 0.286 & 22.4 & N/A & H \\
J1209.4+4340 & NGC4138  & Sy2       & 182.35 & 43.70 & 30.70 & 0.25 & 13.8 & 0.115 & 24.4 & N/A & H \\
J1157.8+5529 & NGC3998  & Sy1.9     & 179.46 & 55.44 & 30.73 & 0.19 & 14.0 & 0.087 & 13.2 & N/A & H \\
J2235.9-2602 & NGC7314  & Sy1.9     & 338.95 & -26.05 & 31.03 & 0.25 & 16.1 & 0.115 & 57.4 & N/A & H \\
J1432.8-4412 & NGC5643  & Sy2       & 218.19 & -44.15 & 31.03 & 1.00 & 16.1 & 0.461 & 16.8 & N/A & H \\
J1001.7+5543 & NGC3079  & Sy2       & 150.46 & 55.67 & 31.16 & 0.32 & 17.1 & 0.147 & 36.7 & N/A & H \\
J1341.9+3537 & NGC5273  & Sy1.5     & 205.47 & 35.66 & 31.16 & 0.12 & 17.1 & 0.055 & 16.0 & N/A & H \\
J1207.8+4311 & NGC4117  & Sy2       & 181.95 & 43.12 & 31.18 & 0.94 & 17.2 & 0.433 & 12.9 & N/A & H \\
J0333.6-3607 & NGC1365  & Sy2       & 53.39 & -36.14 & 31.19 & 0.02 & 17.3 & 0.009 & 63.5 & N/A & H \\
J0241.3-0816 & NGC1052  & BeamedAGN & 40.29 & -8.24 & 31.22 & 0.11 & 17.5 & 0.051 & 31.4 & N/A & H \\
J1132.7+5301 & NGC3718  & Sy1.9     & 173.22 & 53.02 & 31.25 & 0.89 & 17.8 & 0.410 & 12.2 & N/A & H \\
J1206.2+5243 & NGC4102  & Sy2       & 181.59 & 52.71 & 31.29 & 0.25 & 18.1 & 0.115 & 32.1 & N/A & H \\
J2318.4-4223 & NGC7582  & Sy2       & 349.60 & -42.37 & 31.41 & 0.10 & 19.1 & 0.046 & 82.3 & N/A & H \\
\dots & \dots & \dots & \dots & \dots & \dots & \dots & \dots & \dots & \dots & \dots & \dots \\
J0534.8-6026 & 2MASXJ05343093-6016153 & Sy1 & 83.70 & -60.27 & 36.98 & 0.06 & 248.9 & 0.028 & 10.7 & N/A & H
\enddata
\tablecomments{523 entries within 250\:Mpc. 201 entries at $d_{\rm L}< 100\:$Mpc, 458 at $d_{\rm L}< 200\:$Mpc.}
\end{deluxetable}

\end{footnotesize}
\end{longrotatetable}

\begin{longrotatetable}
\begin{footnotesize}

\begin{deluxetable}{cccccccccccc}
\tablecaption{Jetted AGNs (\textit{Fermi}-LAT 3FHL).\label{tab:gagn}}
\tablehead{\colhead{3FHL Name} & \colhead{Counterpart} & \colhead{Jetted AGN Type} & \colhead{R.A.} & \colhead{Dec} & \colhead{$(m-M)$} & \colhead{$\sigma(m-M)$} & \colhead{$d_{\rm L}$} & \colhead{$\sigma(d_{\rm L})/d_{\rm L}$} & \colhead{$\Phi$($0.01-1\:$TeV)} & \colhead{$\sigma(\Phi)$} & \colhead{flag: in Pierre Auger Collaboration 2018b?}\\ \colhead{$\mathrm{}$} & \colhead{$\mathrm{}$} & \colhead{$\mathrm{}$} & \colhead{$\mathrm{{}^{\circ}}$} & \colhead{$\mathrm{{}^{\circ}}$} & \colhead{$\mathrm{mag}$} & \colhead{$\mathrm{mag}$} & \colhead{$\mathrm{Mpc}$} & \colhead{$\mathrm{}$} & \colhead{$10^{-10}\:$cm$^{-2}\:$s$^{-1}$} & \colhead{$10^{-10}\:$cm$^{-2}\:$s$^{-1}$} & \colhead{(No/Yes)}}
\startdata
J1325.5-4300 & CenA & RDG                & 201.37 & -43.02 & 27.83 & 0.03 & 3.68 & 0.014 & 1.54 & 0.25 & Y \\
J1230.8+1223 & M87 & RDG                 & 187.71 & 12.39 & 31.12 & 0.06 & 16.7 & 0.028 & 0.98 & 0.20 & Y \\
J0322.6-3712e & FornaxA & RDG            & 50.67 & -37.21 & 31.55 & 0.03 & 20.4 & 0.014 & 0.48 & 0.16 & N \\
J1346.2-6026 & CenB & RDG                & 206.70 & -60.41 & 33.71 & 0.29 & 55.2 & 0.134 & 0.64 & 0.18 & N \\
J0319.8+4130 & NGC1275 & RDG             & 49.95 & 41.51 & 34.46 & 0.08 & 78.0 & 0.037 & 14.17 & 0.67 & Y \\
J0316.6+4120 & IC310 & RDG               & 49.18 & 41.32 & 34.60 & 0.19 & 83.2 & 0.087 & 0.43 & 0.13 & Y \\
J0153.5+7115 & TXS0149+710 & BCU         & 28.36 & 71.25 & 35.07 & 0.15 & 103.3 & 0.069 & 0.44 & 0.12 & Y \\
J0308.4+0408 & NGC1218 & RDG             & 47.11 & 4.11 & 35.48 & 0.13 & 124.7 & 0.060 & 0.54 & 0.16 & N \\
J1104.4+3812 & Mkn421 & BLL              & 166.10 & 38.21 & 35.63 & 0.12 & 133.7 & 0.055 & 59.35 & 1.38 & Y \\
J1653.8+3945 & Mkn501 & BLL              & 253.47 & 39.76 & 35.91 & 0.10 & 152.1 & 0.046 & 19.17 & 0.76 & Y \\
J0131.1+5546 & TXS0128+554 & BCU         & 22.81 & 55.75 & 36.06 & 0.10 & 162.9 & 0.046 & 0.33 & 0.12 & N \\
J1543.6+0452 & CGCG050-083 & BCU         & 235.89 & 4.87 & 36.26 & 0.09 & 178.6 & 0.041 & 0.69 & 0.17 & N \\
J0223.0-1119 & 1RXSJ022314.6-111741 & BLL & 35.81 & -11.29 & 36.31 & 0.09 & 182.8 & 0.041 & 0.40 & 0.13 & N \\
J2347.0+5142 & 1ES2344+514 & BLL         & 356.76 & 51.69 & 36.47 & 0.08 & 196.8 & 0.037 & 3.32 & 0.31 & Y \\
J0816.4-1311 & PMNJ0816-1311 & BLL       & 124.11 & -13.20 & 36.51 & 0.08 & 200.4 & 0.037 & 2.71 & 0.33 & N \\
J1136.5+7009 & Mkn180 & BLL              & 174.11 & 70.16 & 36.54 & 0.08 & 203.2 & 0.037 & 1.74 & 0.21 & Y \\
J1959.9+6508 & 1ES1959+650 & BLL         & 299.97 & 65.16 & 36.63 & 0.08 & 211.8 & 0.037 & 8.43 & 0.46 & Y \\
J1647.6+4950 & SBS1646+499 & BLL         & 251.90 & 49.83 & 36.64 & 0.08 & 212.8 & 0.037 & 0.48 & 0.12 & N \\
J1517.6-2422 & APLibrae & BLL            & 229.42 & -24.37 & 36.68 & 0.07 & 216.8 & 0.032 & 3.76 & 0.37 & Y \\
J0214.5+5145 & TXS0210+515 & BLL         & 33.55 & 51.77 & 36.70 & 0.11 & 218.8 & 0.051 & 0.42 & 0.12 & Y \\
J1806.8+6950 & 3C371 & BLL               & 271.71 & 69.82 & 36.77 & 0.07 & 225.9 & 0.032 & 1.30 & 0.18 & N \\
J1353.0-4413 & PKS1349-439 & BLL         & 208.24 & -44.21 & 36.79 & 0.07 & 228.0 & 0.032 & 0.33 & 0.12 & N \\
J0200.1-4109 & 1RXSJ020021.0-410936 & BLL & 30.09 & -41.16 & 36.85 & 0.07 & 234.4 & 0.032 & 0.51 & 0.14 & N \\
J0627.1-3528 & PKS0625-35 & BLL          & 96.78 & -35.49 & 36.89 & 0.07 & 238.8 & 0.032 & 1.81 & 0.26 & Y \\
J2039.4+5219 & 1ES2037+521 & BLL         & 309.85 & 52.33 & 36.89 & 0.07 & 238.8 & 0.032 & 0.58 & 0.15 & N \\
J0523.0-3627 & PKS0521-36 & BLL          & 80.76 & -36.46 & 36.91 & 0.07 & 241.0 & 0.032 & 1.17 & 0.21 & N
\enddata
\tablecomments{26 entries within 250\:Mpc. 6 entries at $d_{\rm L}< 100\:$Mpc, 14 at $d_{\rm L}< 200\:$Mpc.}
\end{deluxetable}

\end{footnotesize}
\end{longrotatetable}

\end{changemargin}

\bibliography{biblio}{}
\bibliographystyle{aasjournal}


\newpage

\AuthorCollaborationLimit=3000
{\bf\Large{The Pierre Auger Collaboration}}\\

\input{latex_authorlist_authors.tex}
\input{latex_authorlist_institutions.tex}

\end{document}

%% file: acknowledgments.tex
\section*{Acknowledgments}

\begin{sloppypar}
The successful installation, commissioning, and operation of the Pierre
Auger Observatory would not have been possible without the strong
commitment and effort from the technical and administrative staff in
Malarg\"ue. We are very grateful to the following agencies and
organizations for financial support:
\end{sloppypar}

\begin{sloppypar}
Argentina -- Comisi\'on Nacional de Energ\'\i{}a At\'omica; Agencia Nacional de
Promoci\'on Cient\'\i{}fica y Tecnol\'ogica (ANPCyT); Consejo Nacional de
Investigaciones Cient\'\i{}ficas y T\'ecnicas (CONICET); Gobierno de la
Provincia de Mendoza; Municipalidad de Malarg\"ue; NDM Holdings and Valle
Las Le\~nas; in gratitude for their continuing cooperation over land
access; Australia -- the Australian Research Council; Belgium -- Fonds
de la Recherche Scientifique (FNRS); Research Foundation Flanders (FWO);
Brazil -- Conselho Nacional de Desenvolvimento Cient\'\i{}fico e Tecnol\'ogico
(CNPq); Financiadora de Estudos e Projetos (FINEP); Funda\c{c}\~ao de Amparo \`a
Pesquisa do Estado de Rio de Janeiro (FAPERJ); S\~ao Paulo Research
Foundation (FAPESP) Grants No.~2019/10151-2, No.~2010/07359-6 and
No.~1999/05404-3; Minist\'erio da Ci\^encia, Tecnologia, Inova\c{c}\~oes e
Comunica\c{c}\~oes (MCTIC); Czech Republic -- Grant No.~MSMT CR LTT18004,
LM2015038, LM2018102, CZ.02.1.01/0.0/0.0/16{\textunderscore}013/0001402,
CZ.02.1.01/0.0/0.0/18{\textunderscore}046/0016010 and
CZ.02.1.01/0.0/0.0/17{\textunderscore}049/0008422; France -- Centre de Calcul
IN2P3/CNRS; Centre National de la Recherche Scientifique (CNRS); Conseil
R\'egional Ile-de-France; D\'epartement Physique Nucl\'eaire et Corpusculaire
(PNC-IN2P3/CNRS); D\'epartement Sciences de l'Univers (SDU-INSU/CNRS);
Institut Lagrange de Paris (ILP) Grant No.~LABEX ANR-10-LABX-63 within
the Investissements d'Avenir Programme Grant No.~ANR-11-IDEX-0004-02;
Germany -- Bundesministerium f\"ur Bildung und Forschung (BMBF); Deutsche
Forschungsgemeinschaft (DFG); Finanzministerium Baden-W\"urttemberg;
Helmholtz Alliance for Astroparticle Physics (HAP);
Helmholtz-Gemeinschaft Deutscher Forschungszentren (HGF); Ministerium
f\"ur Innovation, Wissenschaft und Forschung des Landes
Nordrhein-Westfalen; Ministerium f\"ur Wissenschaft, Forschung und Kunst
des Landes Baden-W\"urttemberg; Italy -- Istituto Nazionale di Fisica
Nucleare (INFN); Istituto Nazionale di Astrofisica (INAF); Ministero
dell'Istruzione, dell'Universit\'a e della Ricerca (MIUR); CETEMPS Center
of Excellence; Ministero degli Affari Esteri (MAE); M\'exico -- Consejo
Nacional de Ciencia y Tecnolog\'\i{}a (CONACYT) No.~167733; Universidad
Nacional Aut\'onoma de M\'exico (UNAM); PAPIIT DGAPA-UNAM; The Netherlands
-- Ministry of Education, Culture and Science; Netherlands Organisation
for Scientific Research (NWO); Dutch national e-infrastructure with the
support of SURF Cooperative; Poland -- Ministry of Education and
Science, grant No.~DIR/WK/2018/11; National Science Centre, Grants
No.~2016/22/M/ST9/00198, 2016/23/B/ST9/01635, and 2020/39/B/ST9/01398;
Portugal -- Portuguese national funds and FEDER funds within Programa
Operacional Factores de Competitividade through Funda\c{c}\~ao para a Ci\^encia
e a Tecnologia (COMPETE); Romania -- Ministry of Research, Innovation
and Digitization, CNCS/CCCDI -- UEFISCDI, projects PN19150201/16N/2019,
PN1906010, TE128 and PED289, within PNCDI III; Slovenia -- Slovenian
Research Agency, grants P1-0031, P1-0385, I0-0033, N1-0111; Spain --
Ministerio de Econom\'\i{}a, Industria y Competitividad (FPA2017-85114-P and
PID2019-104676GB-C32), Xunta de Galicia (ED431C 2017/07), Junta de
Andaluc\'\i{}a (SOMM17/6104/UGR, P18-FR-4314) Feder Funds, RENATA Red
Nacional Tem\'atica de Astropart\'\i{}culas (FPA2015-68783-REDT) and Mar\'\i{}a de
Maeztu Unit of Excellence (MDM-2016-0692); USA -- Department of Energy,
Contracts No.~DE-AC02-07CH11359, No.~DE-FR02-04ER41300,
No.~DE-FG02-99ER41107 and No.~DE-SC0011689; National Science Foundation,
Grant No.~0450696; The Grainger Foundation; Marie Curie-IRSES/EPLANET;
European Particle Physics Latin American Network; and UNESCO.
\end{sloppypar}
\begin{sloppypar}
We gratefully acknowledge constructive feedback from the anonymous reviewer as well as exchanges with Alberto Dominguez on the 3FHL catalog and exchanges with Cecilia Lunardini, Kimberly Emig and Rogier Windhorst on their starburst catalog. This research has made use of the SIMBAD database, operated at CDS, Strasbourg, France. This research has made use of the NASA/IPAC Extragalactic Database (NED), which is operated by the Jet Propulsion Laboratory, California Institute of Technology, under contract with the National Aeronautics and Space Administration. We acknowledge the usage of the HyperLeda database (http://leda.univ-lyon1.fr).
\end{sloppypar}

%% file: latex_authorlist_authors.tex
P.~Abreu$^{72}$,
M.~Aglietta$^{54,52}$,
J.M.~Albury$^{13}$,
I.~Allekotte$^{1}$,
K.~Almeida Cheminant$^{70}$,
A.~Almela$^{8,12}$,
J.~Alvarez-Mu\~niz$^{79}$,
R.~Alves Batista$^{80}$,
J.~Ammerman Yebra$^{79}$,
G.A.~Anastasi$^{54,52}$
% ...

%% file: latex_authorlist_institutions.tex
\begin{enumerate}[labelsep=0.2em,align=right,labelwidth=0.7em,labelindent=0em,leftmargin=2em,noitemsep]
\item[$^{1}$] Centro At\'omico Bariloche and Instituto Balseiro (CNEA-UNCuyo-CONICET), San Carlos de Bariloche, Argentina
\item[$^{2}$] Centro de Investigaciones en L\'aseres y Aplicaciones, CITEDEF and CONICET, Villa Martelli, Argentina
\item[$^{3}$] Departamento de F\'\i{}sica and Departamento de Ciencias de la Atm\'osfera y los Oc\'eanos, FCEyN, Universidad de Buenos Aires and CONICET, Buenos Aires, Argentina
\item[$^{4}$] IFLP, Universidad Nacional de La Plata and CONICET, La Plata, Argentina
\item[$^{5}$] Instituto de Astronom\'\i{}a y F\'\i{}sica del Espacio (IAFE, CONICET-UBA), Buenos Aires, Argentina
\item[$^{6}$] Instituto de F\'\i{}sica de Rosario (IFIR) -- CONICET/U.N.R.\ and Facultad de Ciencias Bioqu\'\i{}micas y Farmac\'euticas U.N.R., Rosario, Argentina
\item[$^{7}$] Instituto de Tecnolog\'\i{}as en Detecci\'on y Astropart\'\i{}culas (CNEA, CONICET, UNSAM), and Universidad Tecnol\'ogica Nacional -- Facultad Regional Mendoza (CONICET/CNEA), Mendoza, Argentina
\item[$^{8}$] Instituto de Tecnolog\'\i{}as en Detecci\'on y Astropart\'\i{}culas (CNEA, CONICET, UNSAM), Buenos Aires, Argentina
\item[$^{9}$] International Center of Advanced Studies and Instituto de Ciencias F\'\i{}sicas, ECyT-UNSAM and CONICET, Campus Miguelete -- San Mart\'\i{}n, Buenos Aires, Argentina
\item[$^{10}$] Observatorio Pierre Auger, Malarg\"ue, Argentina
\item[$^{11}$] Observatorio Pierre Auger and Comisi\'on Nacional de Energ\'\i{}a At\'omica, Malarg\"ue, Argentina
\item[$^{12}$] Universidad Tecnol\'ogica Nacional -- Facultad Regional Buenos Aires, Buenos Aires, Argentina
\item[$^{13}$] University of Adelaide, Adelaide, S.A., Australia
\item[$^{14}$] Universit\'e Libre de Bruxelles (ULB), Brussels, Belgium
\item[$^{15}$] Vrije Universiteit Brussels, Brussels, Belgium
\item[$^{16}$] Centro Brasileiro de Pesquisas Fisicas, Rio de Janeiro, RJ, Brazil
\item[$^{17}$] Centro Federal de Educa\c{c}\~ao Tecnol\'ogica Celso Suckow da Fonseca, Nova Friburgo, Brazil
\item[$^{18}$] Instituto Federal de Educa\c{c}\~ao, Ci\^encia e Tecnologia do Rio de Janeiro (IFRJ), Brazil
\item[$^{19}$] Universidade de S\~ao Paulo, Escola de Engenharia de Lorena, Lorena, SP, Brazil
\item[$^{20}$] Universidade de S\~ao Paulo, Instituto de F\'\i{}sica de S\~ao Carlos, S\~ao Carlos, SP, Brazil
\item[$^{21}$] Universidade de S\~ao Paulo, Instituto de F\'\i{}sica, S\~ao Paulo, SP, Brazil
\item[$^{22}$] Universidade Estadual de Campinas, IFGW, Campinas, SP, Brazil
\item[$^{23}$] Universidade Estadual de Feira de Santana, Feira de Santana, Brazil
\item[$^{24}$] Universidade Federal do ABC, Santo Andr\'e, SP, Brazil
\item[$^{25}$] Universidade Federal do Paran\'a, Setor Palotina, Palotina, Brazil
\item[$^{26}$] Universidade Federal do Rio de Janeiro, Instituto de F\'\i{}sica, Rio de Janeiro, RJ, Brazil
\item[$^{27}$] Universidade Federal do Rio de Janeiro (UFRJ), Observat\'orio do Valongo, Rio de Janeiro, RJ, Brazil
\item[$^{28}$] Universidade Federal Fluminense, EEIMVR, Volta Redonda, RJ, Brazil
\item[$^{29}$] Universidad de Medell\'\i{}n, Medell\'\i{}n, Colombia
\item[$^{30}$] Universidad Industrial de Santander, Bucaramanga, Colombia
\item[$^{31}$] Charles University, Faculty of Mathematics and Physics, Institute of Particle and Nuclear Physics, Prague, Czech Republic
\item[$^{32}$] Institute of Physics of the Czech Academy of Sciences, Prague, Czech Republic
\item[$^{33}$] Palacky University, RCPTM, Olomouc, Czech Republic
\item[$^{34}$] CNRS/IN2P3, IJCLab, Universit\'e Paris-Saclay, Orsay, France
\item[$^{35}$] Laboratoire de Physique Nucl\'eaire et de Hautes Energies (LPNHE), Sorbonne Universit\'e, Universit\'e de Paris, CNRS-IN2P3, Paris, France
\item[$^{36}$] Univ.\ Grenoble Alpes, CNRS, Grenoble Institute of Engineering Univ.\ Grenoble Alpes, LPSC-IN2P3, 38000 Grenoble, France
\item[$^{37}$] Universit\'e Paris-Saclay, CNRS/IN2P3, IJCLab, Orsay, France
\item[$^{38}$] Bergische Universit\"at Wuppertal, Department of Physics, Wuppertal, Germany
\item[$^{39}$] Karlsruhe Institute of Technology (KIT), Institute for Experimental Particle Physics, Karlsruhe, Germany
\item[$^{40}$] Karlsruhe Institute of Technology (KIT), Institut f\"ur Prozessdatenverarbeitung und Elektronik, Karlsruhe, Germany
\item[$^{41}$] Karlsruhe Institute of Technology (KIT), Institute for Astroparticle Physics, Karlsruhe, Germany
\item[$^{42}$] RWTH Aachen University, III.\ Physikalisches Institut A, Aachen, Germany
\item[$^{43}$] Universit\"at Hamburg, II.\ Institut f\"ur Theoretische Physik, Hamburg, Germany
\item[$^{44}$] Universit\"at Siegen, Department Physik -- Experimentelle Teilchenphysik, Siegen, Germany
\item[$^{45}$] Gran Sasso Science Institute, L'Aquila, Italy
\item[$^{46}$] INFN Laboratori Nazionali del Gran Sasso, Assergi (L'Aquila), Italy
\item[$^{47}$] INFN, Sezione di Catania, Catania, Italy
\item[$^{48}$] INFN, Sezione di Lecce, Lecce, Italy
\item[$^{49}$] INFN, Sezione di Milano, Milano, Italy
\item[$^{50}$] INFN, Sezione di Napoli, Napoli, Italy
\item[$^{51}$] INFN, Sezione di Roma ``Tor Vergata'', Roma, Italy
\item[$^{52}$] INFN, Sezione di Torino, Torino, Italy
\item[$^{53}$] Istituto di Astrofisica Spaziale e Fisica Cosmica di Palermo (INAF), Palermo, Italy
\item[$^{54}$] Osservatorio Astrofisico di Torino (INAF), Torino, Italy
\item[$^{55}$] Politecnico di Milano, Dipartimento di Scienze e Tecnologie Aerospaziali , Milano, Italy
\item[$^{56}$] Universit\`a del Salento, Dipartimento di Matematica e Fisica ``E.\ De Giorgi'', Lecce, Italy
\item[$^{57}$] Universit\`a dell'Aquila, Dipartimento di Scienze Fisiche e Chimiche, L'Aquila, Italy
\item[$^{58}$] Universit\`a di Catania, Dipartimento di Fisica e Astronomia ``Ettore Majorana``, Catania, Italy
\item[$^{59}$] Universit\`a di Milano, Dipartimento di Fisica, Milano, Italy
\item[$^{60}$] Universit\`a di Napoli ``Federico II'', Dipartimento di Fisica ``Ettore Pancini'', Napoli, Italy
\item[$^{61}$] Universit\`a di Palermo, Dipartimento di Fisica e Chimica ''E.\ Segr\`e'', Palermo, Italy
\item[$^{62}$] Universit\`a di Roma ``Tor Vergata'', Dipartimento di Fisica, Roma, Italy
\item[$^{63}$] Universit\`a Torino, Dipartimento di Fisica, Torino, Italy
\item[$^{64}$] Benem\'erita Universidad Aut\'onoma de Puebla, Puebla, M\'exico
\item[$^{65}$] Unidad Profesional Interdisciplinaria en Ingenier\'\i{}a y Tecnolog\'\i{}as Avanzadas del Instituto Polit\'ecnico Nacional (UPIITA-IPN), M\'exico, D.F., M\'exico
\item[$^{66}$] Universidad Aut\'onoma de Chiapas, Tuxtla Guti\'errez, Chiapas, M\'exico
\item[$^{67}$] Universidad Michoacana de San Nicol\'as de Hidalgo, Morelia, Michoac\'an, M\'exico
\item[$^{68}$] Universidad Nacional Aut\'onoma de M\'exico, M\'exico, D.F., M\'exico
\item[$^{69}$] Universidad Nacional de San Agustin de Arequipa, Facultad de Ciencias Naturales y Formales, Arequipa, Peru
\item[$^{70}$] Institute of Nuclear Physics PAN, Krakow, Poland
\item[$^{71}$] University of \L{}\'od\'z, Faculty of High-Energy Astrophysics,\L{}\'od\'z, Poland
\item[$^{72}$] Laborat\'orio de Instrumenta\c{c}\~ao e F\'\i{}sica Experimental de Part\'\i{}culas -- LIP and Instituto Superior T\'ecnico -- IST, Universidade de Lisboa -- UL, Lisboa, Portugal
\item[$^{73}$] ``Horia Hulubei'' National Institute for Physics and Nuclear Engineering, Bucharest-Magurele, Romania
\item[$^{74}$] Institute of Space Science, Bucharest-Magurele, Romania
\item[$^{75}$] University Politehnica of Bucharest, Bucharest, Romania
\item[$^{76}$] Center for Astrophysics and Cosmology (CAC), University of Nova Gorica, Nova Gorica, Slovenia
\item[$^{77}$] Experimental Particle Physics Department, J.\ Stefan Institute, Ljubljana, Slovenia
\item[$^{78}$] Universidad de Granada and C.A.F.P.E., Granada, Spain
\item[$^{79}$] Instituto Galego de F\'\i{}sica de Altas Enerx\'\i{}as (IGFAE), Universidade de Santiago de Compostela, Santiago de Compostela, Spain
\item[$^{80}$] IMAPP, Radboud University Nijmegen, Nijmegen, The Netherlands
\item[$^{81}$] Nationaal Instituut voor Kernfysica en Hoge Energie Fysica (NIKHEF), Science Park, Amsterdam, The Netherlands
\item[$^{82}$] Stichting Astronomisch Onderzoek in Nederland (ASTRON), Dwingeloo, The Netherlands
\item[$^{83}$] Universiteit van Amsterdam, Faculty of Science, Amsterdam, The Netherlands
\item[$^{84}$] Case Western Reserve University, Cleveland, OH, USA
\item[$^{85}$] Colorado School of Mines, Golden, CO, USA
\item[$^{86}$] Department of Physics and Astronomy, Lehman College, City University of New York, Bronx, NY, USA
\item[$^{87}$] Michigan Technological University, Houghton, MI, USA
\item[$^{88}$] New York University, New York, NY, USA
\item[$^{89}$] Pennsylvania State University, University Park, PA, USA
\item[$^{90}$] University of Chicago, Enrico Fermi Institute, Chicago, IL, USA
\item[$^{91}$] University of Delaware, Department of Physics and Astronomy, Bartol Research Institute, Newark, DE, USA
\item[$^{92}$] University of Wisconsin-Madison, Department of Physics and WIPAC, Madison, WI, USA
\item[] -----
\item[$^{a}$] Fermi National Accelerator Laboratory, Fermilab, Batavia, IL, USA
\item[$^{b}$] Max-Planck-Institut f\"ur Radioastronomie, Bonn, Germany
\item[$^{c}$] School of Physics and Astronomy, University of Leeds, Leeds, United Kingdom
\item[$^{d}$] Colorado State University, Fort Collins, CO, USA
\item[$^{e}$] now at Hakubi Center for Advanced Research and Graduate School of Science, Kyoto University, Kyoto, Japan
\item[$^{f}$] Louisiana State University, Baton Rouge, LA, USA
\item[$^{g}$] also at University of Bucharest, Physics Department, Bucharest, Romania
\end{enumerate}